\documentclass{article}
\usepackage{authblk}
\usepackage{tabularx}
\usepackage[margin=1in]{geometry}
\usepackage[onehalfspacing]{setspace}
\usepackage{amsmath,mathtools}
\usepackage{pgfplots}
\usepackage{subcaption}
\usepackage{booktabs}
\usepackage[braket, qm]{qcircuit}
\usepackage{graphicx}
\usepackage{placeins}
\usepackage{authblk}
 \usepackage{xcolor}
\usepackage{soul}
\usepackage[linesnumbered,ruled,vlined]{algorithm2e}

\SetCommentSty{mycommfont}

\SetKwInput{KwInput}{Input}                
\SetKwInput{KwOutput}{Output}              
\SetKwProg{Fn}{Function}{}{}

\newcommand{\QWalsh}{\mathcal{H}_Q}

\usepackage[english]{babel}

\usepackage{pdflscape}
\usepackage{fancyvrb}
\usepackage{calc}
\usepackage{rotating}
\usepackage{pgf,tikz}
\usepackage{mathrsfs}
\usetikzlibrary{arrows}
\usepackage{mathtools}

\DeclarePairedDelimiter\floor{\lfloor}{\rfloor}
\usepackage{listings}
\usepackage[]{qcircuit}
\usepackage{braket}
\usepackage{mathtools}
\usepackage{varwidth}
\usepackage{caption}
\usepackage{subcaption}
\usepackage{graphicx}
\usepackage[export]{adjustbox}

\usepackage{tcolorbox}

\newcommand{\norm}[1]{\left\lVert#1\right\rVert}

\makeatletter
\def\paragraph{\@startsection{paragraph}{4}%
	\z@\z@{-\fontdimen2\font}%
	{\normalfont\bfseries}}
\makeatother

\usepackage{graphicx}
\usepackage{pgffor}
\usepackage{tikz,tikz-cd}
\usetikzlibrary{backgrounds,fit,decorations.pathreplacing}  
\usepackage{booktabs}
\usepackage{enumerate}

\usepackage{amssymb, amsmath, amsthm}

\usepackage{xypic}
\usepackage{scalerel}
\usepackage{ulem}

\usepackage{graphicx}              
\usepackage{amsmath}               
\usepackage{amsfonts}              
\usepackage{amsthm}                
\usepackage{xkeyval}               
\usepackage{amssymb}
\usepackage{enumerate}             
\usepackage{color}
\usepackage{breqn}                 
\usepackage{bm}                    

\allowdisplaybreaks[1]
\usepackage{nicefrac}
\usepackage{booktabs}

\usepackage{etoolbox}

\usepackage{kpfonts}
\usepackage{stackengine}
\usepackage{calc}
\newlength\shlength
\newcommand\xshlongvec[2][0]{\setlength\shlength{#1pt}%
	\stackengine{-5.6pt}{$#2$}{\smash{$\kern\shlength%
			\stackengine{7.55pt}{$\mathchar"017E$}%
			{\rule{\widthof{$#2$}}{.57pt}\kern.4pt}{O}{r}{F}{F}{L}\kern-\shlength$}}%
	{O}{c}{F}{T}{S}}

\usepackage{hyperref}
\hypersetup{
	colorlinks   = true,    
	citecolor    = red      
}

\newcommand{\RN}[1]{%
	\textup{\uppercase\expandafter{\romannumeral#1}}%
}

\allowdisplaybreaks

\newcommand{\meqref}[1]{\text{Eq}.~\eqref{#1}}
\newcommand{\mref}[1]{Sec.~$ \!\ref{#1} $}
\newcommand{\mfig}[1]{Fig.~$ \!\ref{#1} $}

\newtheorem{thm}{Theorem}[subsection]

\newtheorem{defn}[thm]{Definition} 
\newtheorem{example}[thm]{Example} 
\newtheorem{remark}[thm]{Remark}

\newcommand{\RR}{\mathbb{R}}      

\newcommand{\mat}[4]{\left[\begin{smallmatrix*}[r]
		#1 & #2 \\
		#3 & #4 \\
	\end{smallmatrix*}\right]}

\newcommand{\mseqgset}[3]{%
   \pgfmathparse{int(#1 + #3)}%
  \text{ $ \{ #1$, $\pgfmathresult$, $\ldots \,$, $#2 \}$} %
}

\newcommand{\mseqset}[2]{%
    \pgfmathparse{}%
   \mseqgset{#1}{#2}{1}
}

\newcommand{\mseqsetx}{%
    \pgfmathparse{}%
   \mseqgset{0}{N_1 - 1}{1}
}

\newcommand{\mseqsety}{%
    \pgfmathparse{}%
   \mseqgset{0}{N_2 - 1}{1}
}

\def\NN{{\mathbb N}}

\def\RR{{\mathbb R}}

\def\<{\langle}
\def\>{\rangle}

\newcommand{\uarea}{uniform area measure}
\newcommand{\Uarea}{Uniform area measure}

\newcommand{\uradial}{uniform radial measure}
\newcommand{\Uradial}{Uniform radial measure}

\newcommand{\QWHT}{\mathcal{H}_Q}

\newcommand{\Qwalsh}{\mathcal{H}_Q}

\usetikzlibrary{arrows.meta, quotes}
\makeatletter
\tikzset{
  node on line/.style={
    to path={
      \pgfextra{%
        \edef\tikz@temp{
          edge[
            line to, path only, 
            every edge quotes/.append style={auto=false},
            nodes={alias=@nodeonline@}]
          coordinate(@nodeonline@)
          \unexpanded\expandafter{\tikz@tonodes}(\tikztotarget)
        }\expandafter
      }\tikz@temp
      -- (@nodeonline@) -- (\tikztotarget)}}}
\makeatother

\numberwithin{equation}{section}

\usepackage{bitset}
\pgfplotsset{compat=1.18}

\begin{document}

 \title{Hybrid classical-quantum image processing via polar Walsh basis functions}

	\author[1]{Mohit Rohida}
	\author[2]{Alok Shukla}
	\author[3]{Prakash Vedula}
	\affil[1,2]{School of Arts and Sciences, Ahmedabad University, India}
    \affil[3]{School of Aerospace and Mechanical Engineering, University of Oklahoma, USA}
	\affil[1]{mohit.r@ahduni.edu.in}
	\affil[2]{alok.shukla@ahduni.edu.in}
	\affil[3]{pvedula@ou.edu}

	\maketitle

	\begin{abstract}
 We propose a novel hybrid classical-quantum approach for image processing based on polar Walsh basis functions. Using this approach, we present an algorithm for the removal of the circular banding noise (including Airy pattern noise) and the azimuthal banding noise. This approach is based on a formulation of Walsh basis functions in polar coordinates for image representations. This approach also builds upon an earlier work on a hybrid classical-quantum algorithm for Walsh-Hadamard transforms. We provide two kinds of polar representations using uniform area measure and uniform radial measure. 
Effective smoothening and interpolating techniques are devised relevant to the transformations between Cartesian and polar coordinates, mitigating the challenges posed by the non-injectivity of the transformation in the context of digital images.
The hybrid classical-quantum approach presented here involves an algorithm for Walsh-Hadamard transforms, which has a lower computational complexity of $\mathcal{O}(N)$ compared to the well-known classical Fast Walsh-Hadamard Transform, which has a computational complexity of $\mathcal{O}(N \log_2 N)$.
We demonstrated the applicability of our approach through computational examples involving the removal of the circular banding noise  (including Airy pattern noise) and the azimuthal banding noise. 
\end{abstract}


\section{Introduction}\label{sec:intro}
In recent years, quantum computing algorithms have been proposed in a diverse range of areas, demonstrating superiority over corresponding classical algorithms, including the Deutsch-Jozsa algorithm \cite{deutsch1992rapid}, the Bernstein–Vazirani algorithm \cite{bernstein1993quantum} along with its probabilistic generalization \cite{shukla2023generalization}, Simon's algorithm \cite{simon1997power}, Grover's algorithm \cite{grover1996fast}, Shor's Algorithm \cite{shor1999polynomial}, trajectory optimization \cite{shukla2019trajectory}, solution of linear systems of equations~\cite{harrow2009quantum} , solution of linear and nonlinear differential equations~\cite{shukla2021hybrid,childs2020quantum,berry2014high} and digital signal processing \cite{shukla2023quantum}. 

Several quantum algorithms for image processing have also been  studied~\cite{wang2022review,yao2017quantum,beach2003quantum}. Employing quantum algorithms in image processing requires appropriate quantum image representations \cite{Yan2016ASO}. Frequently used quantum image representations include Real Ket representation, qudit lattices, FRQI, NEQR, and QPIR~\cite{ruan2021quantum,ruan2016quantum}. The quantum image representation and image processing method used in this work is based on the approach introduced in \cite{shukla2022hybrid}. This image representation uses sequential processing of the image subsets (i.e., columns or rows) and requires only $\log_2N$ qubits for an $N \times N$ image, making it efficient compared to its counterparts, such as Real Ket representation, FRQI, and qudit lattice, which require $2\log_2N$, $N^2$, and $2\log_2N+1$ qubits, respectively.

Traditionally, Fourier transform is used for image processing~\cite{moallemi2010adaptive,yaroslavsky1996fundamentals}. Walsh-Hadamard transform~\cite{beauchamp1975walsh} is also used in image processing applications as it can be performed using just additions and subtractions of vector elements,  as opposed to the more computationally expensive trigonometric functions used in Fourier transforms. In quantum image processing applications,  Quantum Fourier Transform (QFT) could be employed.
However, in this work, we will use the Walsh-Hadamard transform. This choice is motivated by the fact that the Walsh-Hadamard transform can be efficiently computed on a quantum computer using Hadamard gates. It is worth noting that for an input vector of size $N=2^n$, QFT has a gate complexity and circuit depth of $\mathcal{O}(n^2)$, whereas the Walsh-Hadamard transform requires only $n$ Hadamard gates and has the circuit depth of $1$. However, the measurement process poses a challenge for both QFT and the Walsh-Hadamard transform, as it allows only the determination of the square of the absolute value of components of the transformed vector. Despite assuming real input sequences, the components of the Walsh-Hadamard transformed vector \( \widehat{\bf{v}} \) may exhibit both positive and negative values, with the sign information being lost during measurement. In many image processing applications, such as grayscale representation where pixel values range from \(0\) to \(255\), applying Walsh-Hadamard transforms to row/column vectors may yield vectors containing negative values, introducing challenges in obtaining unambiguous measurements. As described in \cite{shukla2021hybrid, shukla2022hybrid}, the hybrid classical-quantum approach to image processing applications involves an appropriate adaptation of the quantum Walsh-Hadamard transform and it provides an efficient method for tackling some of these measurement challenges. We note that the classical Fast Walsh-Hadamard Transform~\cite{beauchamp1975walsh} for an input vector of size $ N $ has a computational complexity of order $ \mathcal{O} (N \log_2 N) $, whereas the hybrid classical-quantum algorithm (Algorithm~$ \ref{alg_QWHT} $,  \cite{shukla2021hybrid}) for computation of the Walsh-Hadamard transform for an input vector of size $ N $ is of order $ \mathcal{O}(N) $.

In this work, we propose and discuss two distinct polar representations of Walsh basis functions:  (a)  With a \uarea \, and (b) With a \uradial. In both cases, the circular disc, on which polar Walsh basis functions are supported, is divided into concentric annular regions. For the former case, concentric circles are divided such that each annular sector (or region) has the same area (refer \mfig{area_distribution}). This translates to having roughly the same number of pixels in an image in each annular sector. For the latter case, the radius of each annular region is obtained by dividing the radius of the disc in equal parts (refer \mfig{radial_distribution}). \mref{subsec:polar_Walsh_functions} provides a detailed discussion on polar representations of Walsh basis functions.

Image representation and analysis in polar coordinates are important in many applications where circular symmetry is observed. For example, understanding Airy disks and patterns~\cite{airy1835diffraction,born2013principles}, is important for studying the behavior of light in diverse optical systems and astronomical observations. While the Airy pattern encompasses the entire diffraction pattern, including the central disk and surrounding rings of light, the Airy disk specifically denotes the central bright spot. These phenomena emerge from the wave nature of light interacting with obstacles like circular apertures or point sources. The size of the Airy disk depends on factors like the wavelength of light and the aperture's diameter or the point source's angular size, rendering them indispensable parameters in optical system design and astronomical imaging.

We will apply the polar representations of Walsh basis functions, as discussed earlier, in image processing applications, particularly in cases with circular symmetry. One application involves filtering images affected by circular banding noise, while another focuses on removing noise caused by the Airy Disk and Airy pattern. Computational examples will be provided to illustrate our hybrid classical-quantum approach. This approach relies on Walsh-Hadamard basis functions and their corresponding transforms, in addition to the polar representation of images. Its purpose is to effectively filter noisy images containing circular, Airy Disk, and Airy pattern noises.

Next, we present a brief outline of the paper.
In \mref{sec:wh_transform}, we provide a concise overview of Walsh-Hadamard Transforms. \mref{subsec:2d_wh_transform} describes the two-dimensional Walsh-Hadamard Transforms. In \mref{subsec:polar_Walsh_functions}, we discuss Walsh basis functions in polar coordinates, exploring two distinct polar representations based on (a)  \uarea \, and (b)  \uradial \, and algorithms for obtaining the polar representation from the Cartesian representation of an image and vice versa.  
In  \mref{subsec:polar_representation}, polar image representations using \uarea \, and \uradial \, are discussed.
In \mref{sec:conversion_pc}, we provide algorithms for the conversion of an image from its Cartesian representation to its polar representation by employing the framework described in Section \ref{subsec:polar_representation}. A framework for converting an image from its polar representation to its Cartesian representation is also discussed in \mref{sec:conversion_pc}.
In \mref{sec:removal_circular_noise}, we introduce a hybrid classical-quantum algorithm,  Algorithm \ref{alg_circular_noise_removal}, for removing periodic banding noises.
\mref{sec:computational_example} presents several computational examples to demonstrate the application of Algorithm \ref{alg_circular_noise_removal} in filtering noisy images containing circular (with respect to both \uradial \, and \uarea) banding noise, azimuthal banding noise, and Airy pattern noise.
In Section \ref{sec:disc}, we conclude the article, summarizing the key findings and contributions of this work.

\subsection{Notation} \label{sec:notation} Here we fix some convenient notations used in the rest of the paper.  
	\begin{itemize}
		\item $ \oplus $ : $ x \oplus y $ will denote $ x + y \pmod 2 $.
	    \item $ j \cdot k $ : For $ j = j_{n-1}\,j_{n-2}\,\ldots \, j_1\, j_0 $ and $ k =  k_{n-1}\,k_{n-2} \,\ldots \, k_1\, k_0 $ with $ j_i,\, k_i \in \{0,1\}$, $ j \cdot k $ will denote the bit-wise dot product of  $j $ and $ k $ modulo $ 2 $, i.e.,  $j \cdot k :=  j_{0}k_{0} + j_{1}k_{1}+ \ldots + j_{n-1}k_{n-1} \pmod 2 $.
	\end{itemize}

\section{Walsh-Hadamard transforms}\label{sec:wh_transform}

The Walsh-Hadamard transforms have been widely used in image-processing applications~\cite{yarlagadda2012hadamard}. In this section, we briefly describe Walsh-Hadamard transforms in both natural and sequency orders through their action on computational basis states. More details can be found in references \cite{beauchamp1975walsh, shukla2023quantum}. 

Assume $N=2^n$ is a positive integer. Let $V$ denote the $N$-dimensional complex vector space formed by computational basis states $\{ \ket{0}, \, \ket{1}, \, \ldots \,,\, \ket{N-1} \}$. The Walsh-Hadamard transform in natural order, denoted as $H_N : V \to V$, acts on the computational basis state $\ket{j}$, where $0 \leq j \leq N-1$, according to the equation
\begin{equation} \label{eq_def_Walsh_transform}
    H_N \, \ket{j} = \frac{1}{\sqrt{N}} \sum_{k=0}^{N-1} \, (-1)^{j \cdot k} \, \ket{k}.
\end{equation}
Here, $j \cdot k$ denotes the bit-wise dot product of $j$ and $k$, i.e., if $j = \sum_{i=0}^{N-1} j_i 2^i$ and  $k = \sum_{i=0}^{N-1} k_i 2^i$ then $j\cdot k = \sum_{i=0}^{N-1} j_i k_i$, where $j_i, k_i \in \{0,~1\}$. The corresponding matrix element is given by $(-1)^{j \cdot k}$, resulting in a symmetric matrix $H_N$, known as the Walsh-Hadamard transform matrix in the natural order.
Further, the Walsh-Hadamard transform in sequency order, which we denote by $H^S_N: V \to V$, acts on the basis state $\ket{j}$, with $0 \leq j \leq N-1$, as described by the equation
\begin{equation}\label{eq:sequency:hadamard}
    H_N^S \, \ket{j} = \frac{1}{\sqrt{N}} \sum_{k=0}^{N-1} \, (-1)^{ \sum_{r=0}^{n-1} \, k_{n-1-r} (j_r \oplus j_{r+1}) } \, \ket{k},
\end{equation}
where $k = k_{n-1}\,k_{n-2}\,\ldots \, k_1\, k_0 $ and $j =  j_{n-1}\,j_{n-2} \,\ldots \, j_1\, j_0 $ represent binary representations of $k$ and $j$, with $k_i,\, j_i \in \{0,1\}$ for $ i=0,\,1,\, \ldots ,\,n-1$, and $j_{n} = 0$.

The Walsh-Hadamard transform matrix of order $ N=8 $ in sequency order $H^S_8 $  and natural order $H_8$ are given below.
\begin{align*}
	H_8 =
		\frac{1}{\sqrt{8}} \,
	\begin{pmatrix*}[r]
		1 & 1 & 1 & 1 & 1 & 1 & 1 & 1  \\
		1 & -1 & 1 & -1 & 1 & -1 & 1 & -1  \\
		1 & 1 & -1 & -1 & 1 & 1 & -1 & -1  \\
		1 & -1 & -1 & 1 & 1 & -1 & -1 & 1  \\
		1 & 1 & 1 & 1 & -1 & -1 & -1 & -1  \\
		1 & -1 & 1 & -1 & -1 & 1 & -1 & 1  \\
		1 & 1 & -1 & -1 & -1 & -1 & 1 & 1  \\
		1 & -1 & -1 & 1 & -1 & 1 & 1 & -1  \\
	\end{pmatrix*}
 &  \quad \text{and} \quad 
 H^S_8 = \frac{1}{\sqrt{8}} \,
	\begin{pmatrix*}[r]
		1 & 1 & 1 & 1 & 1 & 1 & 1 & 1  \\
		1 & 1 & 1 & 1 & -1 & -1 & -1 & -1  \\
		1 & 1 & -1 & -1 & -1 & -1 & 1 & 1  \\
		1 & 1 & -1 & -1 & 1 & 1 & -1 & -1  \\
		1 & -1 & -1 & 1 & 1 & -1 & -1 & 1  \\
		1 & -1 & -1 & 1 & -1 & 1 & 1 & -1  \\
		1 & -1 & 1 & -1 & -1 & 1 & -1 & 1  \\
		1 & -1 & 1 & -1 & 1 & -1 & 1 & -1  \\
	\end{pmatrix*}.
\end{align*} 
More details on sequency-ordered matrices can be found in \cite{shukla2024sequency}.

The Walsh-Hadamard transform has a natural connection to Hadamard gates, which are frequently used in quantum computing~\cite{nielsen_chuang_2010}. The Walsh-Hadamard transform of a normalized vector \( \mathbf{v} \in \mathbb{C}^{N} \) consisting of \( N = 2^{n} \) components (\( n \in \mathbb{N} \)), represented as \( \mathbf{v} = {[f_0 ~~~ f_1 ~~~ f_2 ~~~ \ldots ~~~ f_{N-1}]}^{T} \), can be computed by preparing an \( n \)-qubit state \( \ket{\psi} = \sum_{k=0}^{N-1} f(k) \ket{k} \) and then applying \( H^{\otimes n} \) on this state. If one disregards the state preparation costs (for instance, when the input quantum state results from the partial computation of some other quantum circuit at an intermediate step), the gate complexity of performing the quantum Walsh-Hadamard transform is \( \mathcal{O}(\log_2 N) \) with a circuit depth of \( \mathcal{O}(1) \). This is superior to the classical Fast Walsh-Hadamard transform algorithm, which has a computational complexity of \( \mathcal{O}(N \log_2 N) \). 

Our hybrid classical-quantum approach to image processing applications using polar Walsh representation is based on an adaptation of the quantum Walsh-Hadamard transform. However, the challenge lies in measurement, where one can only determine the square of the amplitudes for the Walsh-Hadamard transformed vector. Since the input sequence is assumed to be real, the components of the Walsh-Hadamard transformed vector \( \widehat{\bf{v}} \) are also real. However, the components of \( \, \widehat{\bf{v}} \, \) may be positive or negative, and this sign information is lost during measurement. It's worth noting that in many image processing applications, pixel values are non-negative. For instance, in grayscale representation, pixel values may range from \( 0 \) to \( 255 \). However, applying the Walsh-Hadamard transforms to row/column vectors may yield vectors containing negative values. This introduces potential challenges in obtaining unambiguous measurements, as mentioned earlier. 
		
The core problem of obtaining Walsh-Hadamard transforms with the correct sign information, by exploiting the structure of the Walsh-Hadamard transform matrix, was addressed in Ref.~\cite{shukla2021hybrid}. 
The approach in Ref. \cite{shukla2021hybrid} depended upon a key lemma (see Lemma 4.0.1 in \cite{shukla2021hybrid}). This resulted in an algorithm of $ \mathcal{O}(N) $ (See Algorithm 1 in \cite{shukla2021hybrid}) to compute the  Walsh-Hadamard transform of an input vector of size $ N $. We reproduce this algorithm below for easy reference.

\begin{algorithm}[H] \label{alg_QWHT}
			\DontPrintSemicolon
			\KwInput{The input vector $ A = [a_0 \quad a_1 \quad a_2 \quad  \ldots \quad a_{N-1} ]^{T} $ where $ N =2^n $ is a positive integer and $ a_i \in \RR $ for $ i=0 $ to $ i=N-1 $. }
			\KwOutput{The Walsh-Hadamard transform  (in the sequency order) of the input vector.}
			\Fn{$ \QWHT $ (A)}{
						$ b_0 = \epsilon + \sum_{k=0}^{N-1} \, |a_k| $ \tcp*{Here $ \epsilon $ is any positive number.}
					$   c = \sqrt{\left[ b_0^2 + \sum_{k=1}^{N-1} a_k^2 \right]}$ \tcp*{ Let $ \widetilde{A} = [b_0 \quad a_1 \quad a_2, \ldots \quad a_{N-1} ]^{T}$. Then $ c = \norm{\widetilde{A}} $. } 
					Prepare the state $ \ket{\Psi} = \frac{b_0}{c} \ket{0} + \sum_{k=1}^{N -1}\, \frac{a_k}{c} \ket{k}$ using $ n $ qubits. \tcp*{Initialize the state $ \ket{\Psi}  $ with $ \frac{\widetilde{A}}{\norm{\tilde{A}}}$.}
					Apply $ H^{\otimes n} $ on $ \ket{\Psi} $. \\
					Measure all the $ n $ qubits to compute the probability $ p_k $ of obtaining the state $ \ket{k} $, for $ k=0 $ to $ 2^n-1 $. \\
					$ \delta = \frac{1}{\sqrt{N}}(b_0 - a_0 )$ \\
					$ {\bf{ u}} =   [c\sqrt{p_0} - \delta \quad c\sqrt{p_1} - \delta \quad c\sqrt{p_2} - \delta \quad \ldots \quad c\sqrt{p_{N-1}} - \delta ]^{T} $ \\
					Convert ${\bf{ u}}$ in the sequency order and store it in the vector ${ \bf{ v}} $. \\
					\Return{ the vector ${ \bf{ v}} $.}
			}
			\caption{A hybrid classical-quantum algorithm for computing the Walsh-Hadamard transform $ \QWHT(A) $ (in the sequency order) of a given input vector $ A $.}
		\end{algorithm}
We note that the parameter $ \epsilon $ ensures that Algorithm~$ \ref{alg_QWHT} $ also works for the special case when the $ \norm{A} =0 $. 
We already noted that the computational complexity of the classical Fast Walsh-Hadamard Transform~\cite{geadah1977natural} for an input vector of size $ N $ is $ \mathcal{O} (N \log_2 N) $, whereas our hybrid classical-quantum algorithm (Algorithm~$ \ref{alg_QWHT} $) for computation of the Walsh-Hadamard transform for an input vector of size $ N $ has a computational complexity of $ \mathcal{O}(N) $.

We observe that sequency order is generally preferred for image processing applications due to its better energy compaction properties. A complete quantum circuit to obtain the Walsh-Hadamard transform in sequency order from natural order can be found in \cite{shukla2022quantum, shukla2023quantum}.

\subsection{Two-dimensional Walsh-Hadamard transform}\label{subsec:2d_wh_transform}
A two-dimensional Walsh-Hadamard transform for an \(N \times N\) matrix \(F\) is the matrix \(\widehat{F}\) of the same size, where for \(0 \leq p,q \leq N-1\), the matrix element \(\widehat{F}_{p,q}\) of \(\widehat{F}\) is  

\begin{equation}\label{eq_two_D_Walsh_transform}
	\widehat{F}_{p,q} = \frac{1}{N} \sum_{r=0}^{N-1} \sum_{s=0}^{N-1} \, (-1)^{p\cdot r + q\cdot s}\,F_{r,s}.
\end{equation}
Equation \eqref{eq_two_D_Walsh_transform} can be re-written as
\begin{equation}\label{eq_two_D_Walsh_transform-2}
	\widehat{F}_{p,q} = \frac{1}{N} \sum_{r=0}^{N-1} \left(\sum_{s=0}^{N-1} (-1)^{q\cdot s} F_{r,s} \right) (-1)^{p\cdot r}.
\end{equation}
Recall that the notation \(p\cdot r\) represents the bitwise dot product between \(p\) and \(r\). An inverse of the two-dimensional Walsh-Hadamard transform for the matrix \(\widehat{F}\) is then similarly defined as
\begin{equation}\label{eq_two_D_Inverse_Walsh_transform}
	F_{r,s} = \frac{1}{N} \sum_{p=0}^{N-1} \left(\sum_{q=0}^{N-1}(-1)^{q\cdot s}\widehat{F}_{p,q} \right) (-1)^{p\cdot r}. 
\end{equation}
where, $0\leq r, s \leq N-1$.

From \meqref{eq_two_D_Walsh_transform-2}, it can be seen that the two-dimensional Walsh-Hadamard transform can be performed by two successive one-dimensional Walsh-Hadamard transforms. More explicitly, for a two-dimensional Walsh-Hadamard transform of a $N\times N$ matrix $F$, first step would be to compute a one-dimensional Walsh-Hadamard transform for each column in the matrix, followed by a one-dimensional Walsh-Hadamard transform for each row in the matrix. 
\vspace{0.23cm}
\normalem 

\begin{algorithm}[H] \label{alg_two_dimensional_Walsh_Transfrom}
\DontPrintSemicolon
\KwInput{A $ N \times N $ matrix $ X $. Here $ N=2^n $ for some positive integer $ n $. }
\KwOutput{The two-dimensional Walsh-Hadamard transform of $ X $.}
\tcc{The algorithm uses the quantum subroutine $ \QWalsh $($ \bf{v} $) to compute the quantum Walsh-Hadamard transform of the input vector $ \bf{v} $ of size $ n $. }
\Fn{$ \QWalsh^{\otimes 2} $ (X)}
{
    \For{$ j \gets 1$ \KwTo $ N$ }
    {
        $ X[j] = \QWalsh (X[j]) $	\tcp*{Replace the $ j^{\text{th}} $ column of $ X $ with its Walsh-Hadamard transform.} 
	}
	\For{$ i \gets 1$ \KwTo $ N$ }
	{
		$ X^T[i] = \QWalsh (X^T[i]) $	\tcp*{Replace the $ i^{\text{th}}$ row of $ X $ with its Walsh-Hadamard transform.} 
	}
    \Return{$ X $.}
}
\caption{A hybrid classical-quantum algorithm for computing two-dimensional Walsh-Hadamard transform.}
		\end{algorithm}
\ULforem 
\vspace{0.25cm}

Based on the approach discussed above, Algorithm \ref{alg_two_dimensional_Walsh_Transfrom} computes a two-dimensional Walsh-Hadamard transform for a $N \times N$ matrix consisting of all real elements, where $N = 2^n,~n\in \NN$. This algorithm is reproduced from \cite{shukla2022hybrid}.
Algorithm \ref{alg_two_dimensional_Walsh_Transfrom} uses a quantum subroutine $\QWalsh(\bf{v})$ that computes the one-dimensional Walsh-Hadamard transform for a vector $\bf{v}~\in \mathbb{R}$ with a computational complexity of $\mathcal{O}(N)$. A detailed discussion for computing a one-dimensional Walsh-Hadamard transform using this approach is provided in \cite{shukla2021hybrid}. Using this method for performing a one-dimensional Walsh-Hadamard transform, an approach to obtain a two-dimensional Walsh-Hadamard transform in Algorithm \ref{alg_two_dimensional_Walsh_Transfrom} would have a computational complexity of $\mathcal{O}(N_1 N_2)$ for a $N_1 \times N_2$ matrix, where $N_1$ and $N_2$ are integer powers of $2$. In contrast, the classical Fast Walsh-Hadamard transform has a computational complexity of $\mathcal{O}(N_1N_2 \log_2(N_1N_2))$ for a similar $N_1\times N_2$ matrix. Similarly, the inverse two-dimensional Walsh-Hadamard transform could be computed by two successive one-dimensional inverse Walsh-Hadamard transforms.
   
\section{Walsh basis functions in polar coordinates}\label{subsec:polar_Walsh_functions}

In this section,  Walsh basis functions in polar coordinates are discussed. A discussion on the set of Walsh basis functions in Cartesian coordinates can be found in \cite{shukla2021hybrid}.

In the following, we will consider two distinct polar representations of Walsh basis functions:  (a)  polar Walsh basis functions with a \uarea \, and (b) polar Walsh basis functions with a \uradial. 
Let $r_{max}$ be the radius of the disk on which the polar Walsh basis functions are supported (i.e., outside of this disk, the polar Walsh basis functions vanish).
Assume $N_1 = 2^{n_1}$ and $N_2 = 2^{n_2}$, where $n_1, n_2 \in \mathbb{N}$.
Polar Walsh basis functions of order $(N_1, N_2)$ are defined by dividing the radius $r_{max}$ into $N_1$ parts and the angle $2 \pi$ into $N_2$ parts as described below.
Let
\begin{equation}
    \theta_{q} =  \frac{2 \pi (1+q) }{N_2},\label{eq_theta_k}
\end{equation} 
for $q=0$ to $q=N_2 - 1$.
Further, for $k=0$ to $N_1-1$ we define the radius of the $k$th concentric circle 
\begin{equation}\label{area_and_radial_distribution}
    r_k = \left(\,\frac{1+k}{N_1}\,\right)^{f}r_{max}, 
\end{equation}
where $f=\frac{1}{2}$ for \uarea \, and $f=1$ for \uradial.

We note that, in \uradial, a disk of radius $r_{max}$ contains concentric annular regions, whose boundaries are formed by $N_1$ concentric circles.  These circles are drawn uniformly, with the radius of the $k$th circle calculated as $r_k = \left(\frac{1+k}{N_1}\right) r_{max}$, for $k=0$ to $N_1-1$. This explains $f = 1$ case in \meqref{area_and_radial_distribution}. In \uarea, a disk of radius $r_{max}$ is divided into $N_1 $  concentric annular regions, such that each annular region has the same area (refer \mfig{area_distribution} for an example for $N_1=4$). It follows that
\begin{equation*}
    \pi r_0^{2} = \pi (r_1^{2}-r_0^{2}) = \pi (r_2^{2}-r_1^{2}) =\ldots= \pi (r_{N_1-1}^{2}-r_{N_1-2}^{2}).
\end{equation*}
Therefore the radius of $k^{th}$- circle is 
\begin{equation}\label{Walsh_radial_distribution}
    r_k = \left(\,\frac{1+k}{N_1}\,\right)^{1/2}r_{max}.
\end{equation}
This explains the $f=\frac{1}{2}$ case in \meqref{area_and_radial_distribution}.
We note that in image processing,  \uarea \, may find broader applications, because each annular region would contain approximately the same number of pixels (as each annular region has the same area). 
One can also obtain other uniform area measures using appropriate partitions of azimuthal (i.e., $\theta$) coordinates. Such uniform area measures will not be considered here.
\begin{example}
In \mfig{area_radial_distribution}, examples of \uarea \, (on the left) and \uradial \, (on the right) are described, respectively, for $N_1 = 4$ divisions of the radius $r_{max}$.
\end{example}

\newcommand{\colorA}{green!20}
\newcommand{\colorB}{orange!20}
\newcommand{\colorC}{red!20}
\newcommand{\colorD}{blue!20}

\begin{figure}[ht]
    \begin{subfigure}{0.45\textwidth}
    \centering
    \begin{tikzpicture}[scale=0.85]
    \pgfmathsetmacro{\ra}{2.0}
    \pgfmathsetmacro{\rb}{2.82}
    \pgfmathsetmacro{\rc}{3.46}
    \pgfmathsetmacro{\rd}{4}
        \fill [\colorD] (0,0) circle [radius=\rd];
        \fill [\colorC] (0,0) circle [radius=\rc];
        \fill [\colorB] (0,0) circle [radius=\rb];
        \fill [\colorA] (0,0) circle [radius=\ra];
        \draw[->|] (0,0) to[node on line, "$r_0$"] (0:\ra);
        \draw[->|, rotate around={30:(0,0)}] 
            (0,0) to[node on line] node[midway]{$r_1$} (\rb,0);
        \draw[->|, rotate around={60:(0,0)}] 
            (0,0) to[node on line] node[midway]{$r_2$} (\rc,0);
        \draw[->|, rotate around={90:(0,0)}] 
            (0,0) to[node on line] node[above]{$r_3$} (\rd,0);
        \draw[fill=black](0,0) circle [radius=\rb pt]
                       node [anchor=south east] {$O$};
    \end{tikzpicture}
    \caption{ \Uarea. }
    \label{area_distribution}
\end{subfigure}
\hfill
\begin{subfigure}{0.45\textwidth}
    \centering
     \begin{tikzpicture}[scale=0.85]
    \pgfmathsetmacro{\ra}{1}
    \pgfmathsetmacro{\rb}{2}
    \pgfmathsetmacro{\rc}{3}
    \pgfmathsetmacro{\rd}{4}
        \fill [\colorD] (0,0) circle [radius=\rd];
        \fill [\colorC] (0,0) circle [radius=\rc];
        \fill [\colorB] (0,0) circle [radius=\rb];
        \fill [\colorA] (0,0) circle [radius=\ra];
        \draw[->|] (0,0) to[node on line, "\footnotesize{$r_0$}"] (0:\ra);
        \draw[->|, rotate around={30:(0,0)}] 
            (0,0) to[node on line] node[midway]{$r_1$} (\rb,0);
        \draw[->|, rotate around={60:(0,0)}] 
            (0,0) to[node on line] node[midway]{$r_2$} (\rc,0);
        \draw[->|, rotate around={90:(0,0)}] 
            (0,0) to[node on line] node[above]{$r_3$} (\rd,0);
        \draw[fill=black](0,0) circle [radius=\rb pt]
                       node [anchor=south east] {$O$};  
    \end{tikzpicture}
  \caption{\Uradial.}
  \label{radial_distribution}
\end{subfigure}
\caption{Examples: (a) \Uarea \,  and (b) \Uradial \, for   $N_1=4$. Note that in the left figure corresponding to \uarea, the areas of the annular regions are equal, i.e. $\pi r_0^2 = \pi (r_1^2 - r_0^2) = \pi (r_2^2 - r_1^2) = \pi (r_3^2 - r_2^2)$. Whereas, in the right figure corresponding to \uradial, we have $r_0 = (r_1-r_0) = (r_2-r_1) = (r_3-r_2)$.}
\label{area_radial_distribution}
\end{figure}
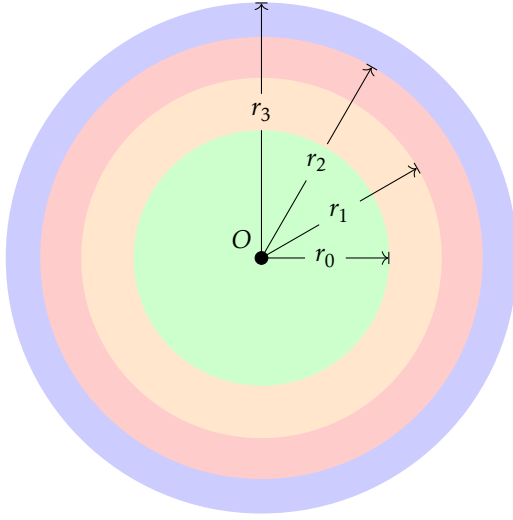
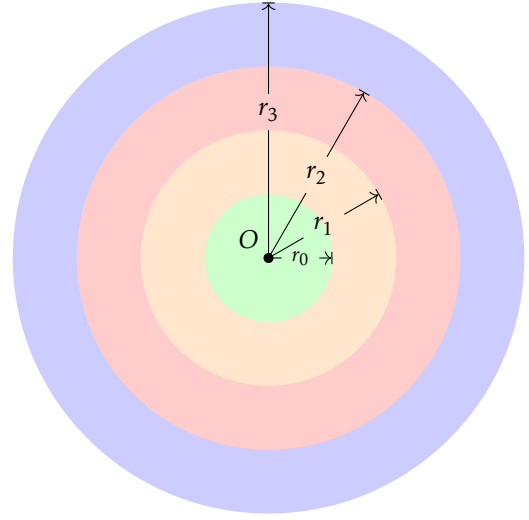

As noted earlier, in addition to natural order, Walsh basis functions can be expressed in sequency order,  which is especially suited for image processing applications. More specific details on the concept of sequency, Walsh functions in sequency order and their applications in image and signal processing can be found in \cite{shukla2023quantum, shukla2022hybrid, beauchamp1975walsh, shukla2024sequency}.

\begin{figure}
  \begin{subfigure}{0.45\textwidth}
    \centering
    \begin{tikzpicture}
        \draw[fill=\colorB] (0,0) -- (3,0) arc (0:180:3) -- cycle;
         \draw[fill=\colorC] (0,0) -- (-3,0) arc (180:360:3) -- cycle;
        \draw[fill=\colorA] (0,0) -- (2.121320,0) arc (0:180:2.121320) -- cycle;
         \draw[fill=\colorD] (0,0) -- (-2.121320,0) arc (180:360:2.121320) -- cycle;
        \fill (0cm,0.5cm) node[black]{$S(0,0)$};
        \fill (0cm,-0.5cm) node[black]{$S(0,1)$};
        \fill (0cm,2.5cm) node[black]{$S(1,0)$};
        \fill (0cm,-2.5cm) node[black]{$S(1,1)$};
    \end{tikzpicture}
    \caption{\Uarea, for $N_1 = 2$, $N_2 = 2$.}
  \end{subfigure}
  \hfill
  \begin{subfigure}{0.45\textwidth}
    \centering
 \begin{tikzpicture}
        \draw[fill=\colorB] (0,0) -- (3,0) arc (0:180:3) -- cycle;
         \draw[fill=\colorC] (0,0) -- (-3,0) arc (180:360:3) -- cycle;
        \draw[fill=\colorA] (0,0) -- (1.5,0) arc (0:180:1.5) -- cycle;
         \draw[fill=\colorD] (0,0) -- (-1.5,0) arc (180:360:1.5) -- cycle;
        \fill (0cm,0.5cm) node[black]{$S(0,0)$};
        \fill (0cm,-0.5cm) node[black]{$S(0,1)$};
        \fill (0cm,2.5cm) node[black]{$S(1,0)$};
        \fill (0cm,-2.5cm) node[black]{$S(1,1)$};
    \end{tikzpicture}

     \caption{\Uradial, for $N_1 = 2$, $N_2 = 2$.}\label{b}
  \end{subfigure}
\caption{Annular  regions or sectors $S(0,0),~S(0,1),~S(1,0)$ and $S(1,1)$ are constructed by partitioning the disk $D$ with $N_1 = 2$ and $N_2 = 2$ for (a) \uarea \, and (b) \uradial.}\label{polar_discretization} 
\end{figure}
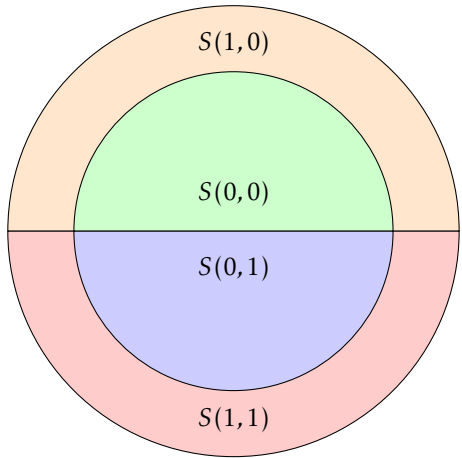
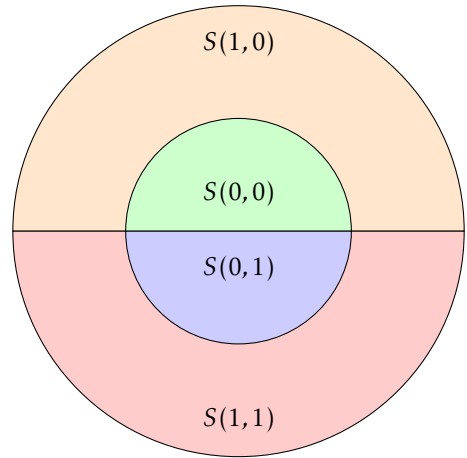

In the following, we provide definitions of Polar Walsh basis functions in both natural order and sequency order in two dimensions. For notational convenience, we define the set $S(k,q)$ as
\begin{equation} \label{eq:skq}
    S(k,q) := \{(r,\theta)  \ | \ r_{k-1} \leq r < r_k,~  \theta_{q-1} \leq \theta < \theta_q\},
\end{equation}
where $k \in \{0, 1, \ldots, N_1 - 1\}$ and $q \in \{0, 1, \ldots, N_2 - 1\}$. Here, $\theta_k$ and $r_k$ are defined as in \meqref{eq_theta_k} and \meqref{area_and_radial_distribution}, respectively. Additionally, we set $r_{-1} = \theta_{-1} = k_{n_1} = q_{n_2} = 0$. 
Clearly, the open disk $D$ of radius $r_{max}$ is partitioned into $N_1 \times N_2$ annular regions or sectors, such that
\begin{equation}\label{eq_def_D}
    D:= \{(x,y) \in \RR^2  \ | \ x^2 + y^2 < r_{max}\} = \bigcup_{(i, j) \in T} \,     S(i,j),
\end{equation}
where $T = \mseqset{0}{N_1 -1}  \times  \mseqset{0}{N_2 -1}$ and the set $S(i,j)$ represents the annular region or the sector indexed by $(i,j)$. This is illustrated in  \mfig{polar_discretization} for $N_1=2$ and $N_2=2$, where the disk $D$ is partitioned into four different annular regions or sectors
represented by $S(0,0)$, $S(0,1)$, $S(1,0)$ and $S(1,1)$.\\

\begin{defn}[\textbf{Polar Walsh basis functions in two dimensions}]
Let $N_1 = 2^{n_1}$ and $N_2 = 2^{n_2}$, where $n_1, n_2 \in \mathbb{N}$.
      For $0 \leq j \leq N_1 - 1$, $0 \leq p \leq N_2 - 1$, $0 \leq r < r_{max}$, and $0 \leq  \theta < 2 \pi$, with $(r,\theta) \in S(k,q)$, the polar Walsh basis function $W_{j,p}(r,\theta)$ of order $(N_1,N_2)$ in natural order is defined as:
\begin{equation}\label{wk_full_polar}
    W_{j,p}(r,\theta) = (-1)^{j \cdot k + p\cdot q}, \quad 
\end{equation} 
and 
the polar Walsh basis function $W_{j,p}^{s}(r,\theta)$ of order $(N_1,N_2)$ in sequency order is defined as:
\begin{equation}\label{wk_full_polar_seq}
    W_{j,p}^s(r,\theta) = (-1)^{ \sum_{i=0}^{n_1-1} \, j_{n_1-1-i} (k_i \oplus k_{i+1})  + 
    \sum_{t=0}^{n_2-1} \, p_{n_2-1-t} (q_t \oplus q_{t+1}) }.\quad 
\end{equation}
We note that the notation $j \cdot k$ refers to the bitwise dot product of $j$ and $k$ (refer \mref{sec:notation}). We further observe that the above definition of $ W_{j,p}(r,\theta)$ is based on \meqref{eq_two_D_Walsh_transform}.

\end{defn}
It is obvious from the above definition of polar Walsh basis functions that they are constant on each annular region $S({k,q})$ for $k \in \mseqsetx$ and $q \in \mseqsety$.  

\begin{figure}
\begin{subfigure}{0.49\textwidth}
        \centering
        \includegraphics[scale=0.4]{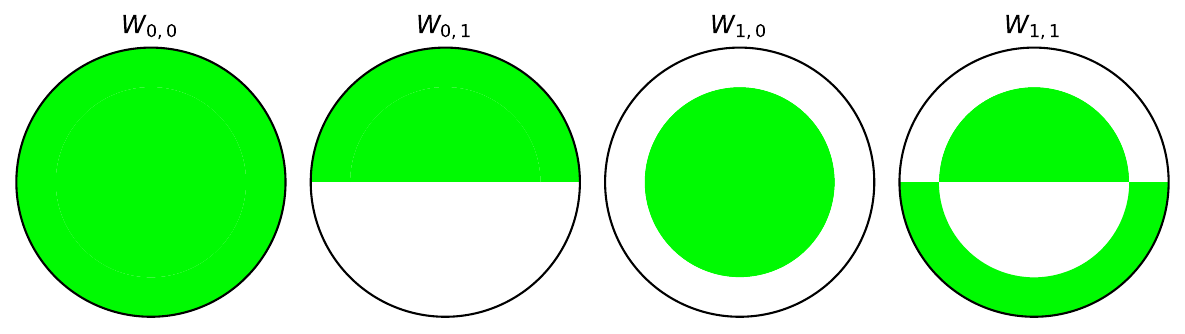}
        \caption{\Uarea, $N_1=2$, $N_2=2$.}
        \label{fig:sub1}
    \end{subfigure}
     \hfill
    \begin{subfigure}{0.49\textwidth}
        \centering
          \includegraphics[scale=0.4]{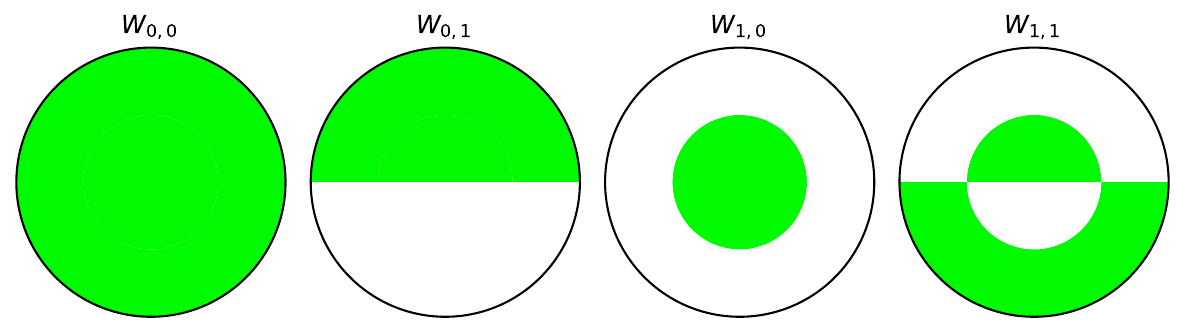}
         \caption{\Uradial, $N_1=2$, $N_2=2$.}
        \label{fig:sub2}
    \end{subfigure}
\vskip 10pt
    \begin{subfigure}{0.49\textwidth}
        \centering
        \includegraphics[scale=0.46]{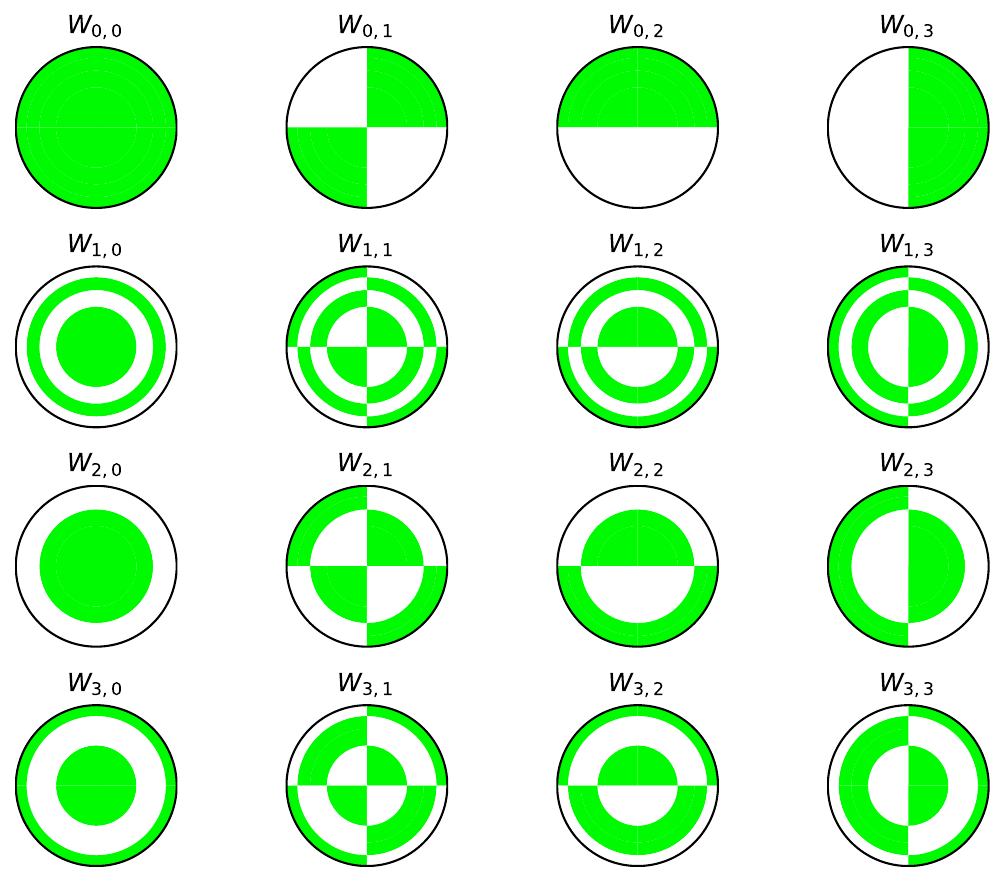}
        \caption{\Uarea, $N_1=4$, $N_2=4$.}
        \label{fig:sub3}
    \end{subfigure}
     \hfill
    \begin{subfigure}{0.49\textwidth}
        \centering
          \includegraphics[scale=0.46]{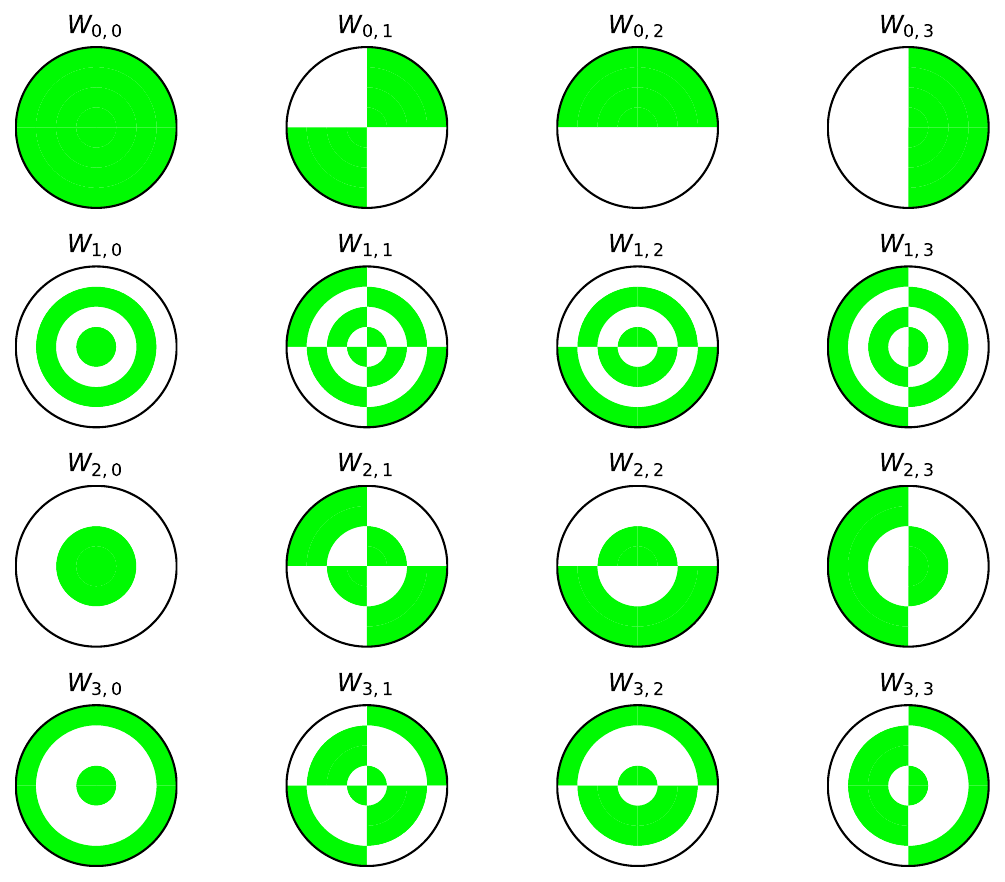}
        \caption{\Uradial, $N_1=4$, $N_2=4$.}
        \label{fig:sub4}
    \end{subfigure}
    \vskip 10pt
   \caption{\Uarea \, and \uradial \, in natural order.} \label{nat_ua_ur_figures}
\end{figure}

\begin{figure}
\begin{subfigure}{0.49\textwidth}
        \centering
        \includegraphics[scale=0.4]{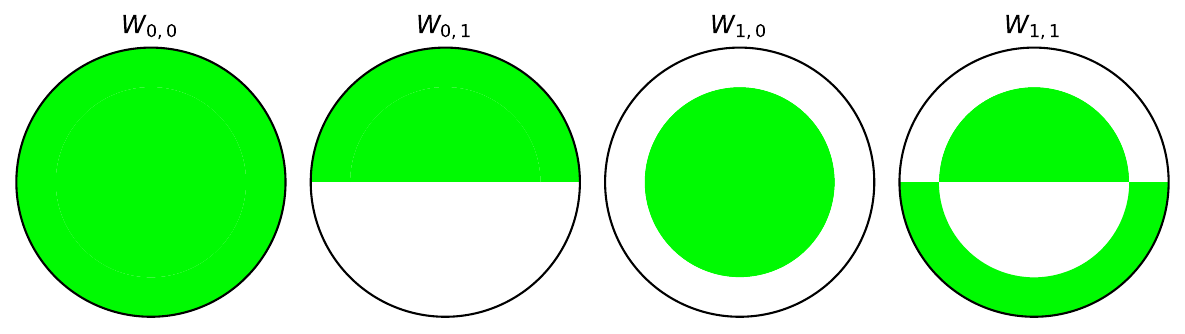}
        \caption{\Uarea, $N_1=2$, $N_2=2$.}
        \label{fig:sub5}
    \end{subfigure}
     \hfill
    \begin{subfigure}{0.49\textwidth}
        \centering
          \includegraphics[scale=0.4]{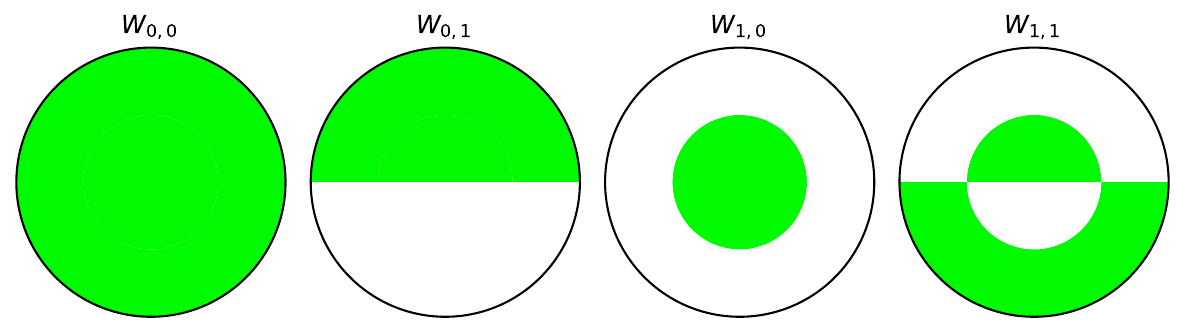}
         \caption{\Uradial, $N_1=2$, $N_2=2$.}
        \label{fig:sub6}
    \end{subfigure}
\vskip 10pt
    \begin{subfigure}{0.49\textwidth}
        \centering
        \includegraphics[scale=0.46]{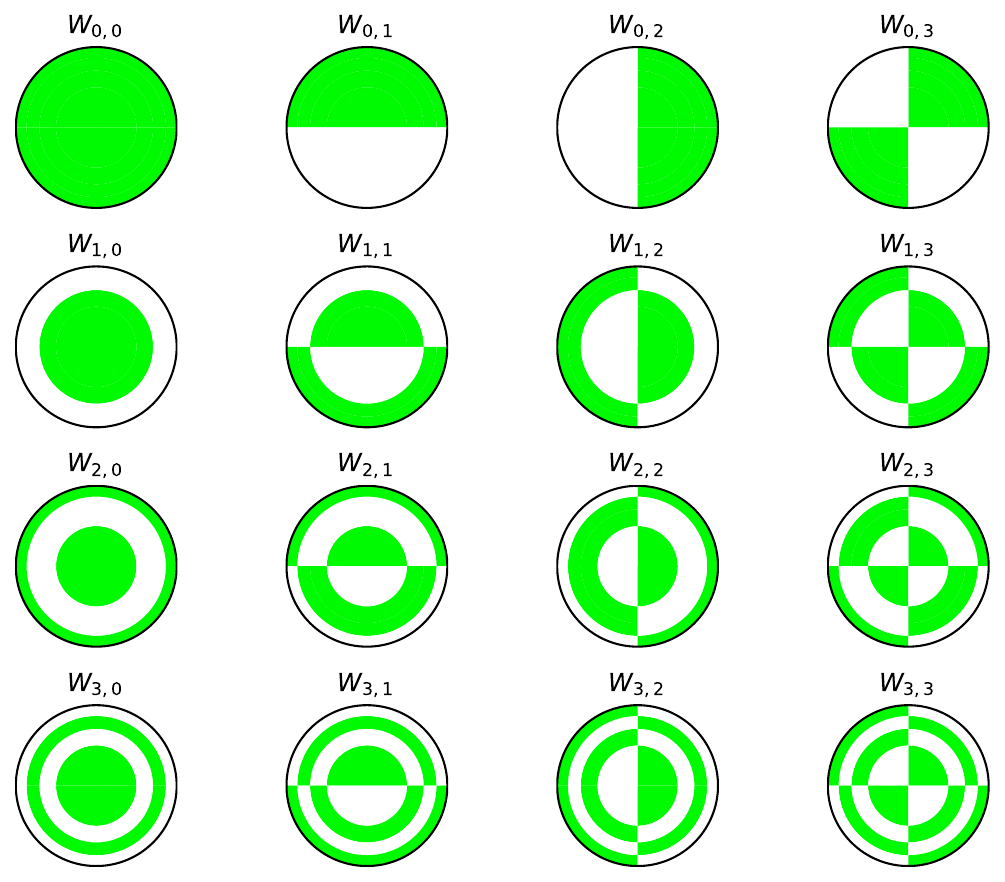}
        \caption{\Uarea, $N_1=4$, $N_2=4$.}
        \label{fig:sub44}
    \end{subfigure}
     \hfill
    \begin{subfigure}{0.49\textwidth}
        \centering
          \includegraphics[scale=0.46]{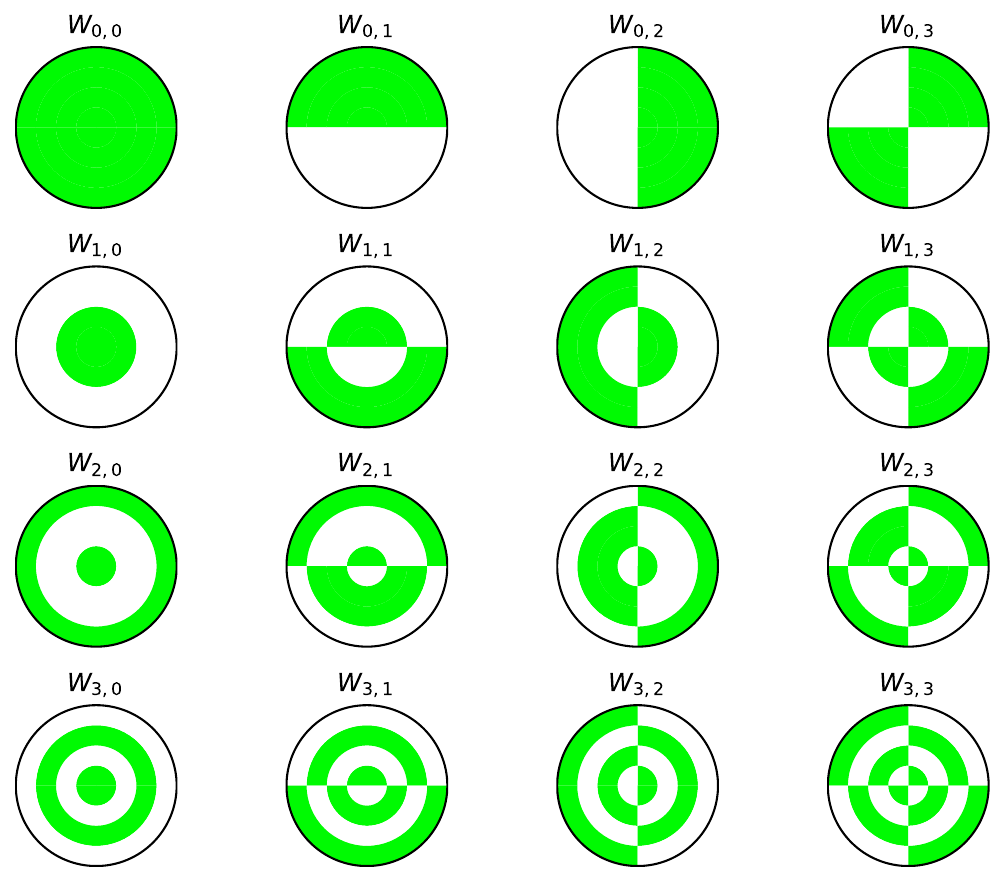}
        \caption{\Uradial, $N_1=4$, $N_2=4$.}
        \label{fig:sub7}
    \end{subfigure}
    \vskip 10pt
   \caption{\Uarea \, and \uradial \, in sequency order.} \label{seq_ua_ur_figures}
\end{figure}

\begin{example} 
Examples of  Polar Walsh basis functions with \uradial \, and \uarea \, for $N_1 = 2$, $N_2 = 2$ and $N_1 = 4$, $N_2 = 4$, in natural order and sequency order  are shown in \mfig{nat_ua_ur_figures} and \mfig{seq_ua_ur_figures}, respectively.
In these figures the color green denotes  $1$, and the color white  $-1$. 
\end{example}


\subsection{Polar image representations}\label{subsec:polar_representation}
In this section, we consider polar image representations using \uarea \ and \uradial. Let the gray-scale input image $\widetilde{I}$ be given as a matrix of size $\widetilde{N}_1 \times \widetilde{N}_2 $ (with $1~{<}~ \widetilde{N}_k \in \mathbb{N}$, for $k=1, \, 2$), such that $\widetilde{I}(i,j)$ denotes the pixel intensity at the location $(i,j)$, where $(i,j)$ is the Cartesian coordinate of the pixel location. 
One can think of the image $\widetilde{I}$ as a function  
\[
\widetilde{I} : D_x \times D_y \longrightarrow G,
\]
where $D_x$ and $D_y$ are discrete sets of size $\widetilde{N}_1 $ and $\widetilde{N}_2$, respectively, and  $G$ is the set of values for pixel intensities. For example, for a gray-scale image $\widetilde{I}$ of size $512 \times 256$, one can have $D_x = \{ 0,\, \ldots,, \, 511\}$, $D_y = \{0,\, 1,\, \ldots,\,255\}$, and  $G = \{0,\, 1,\, \ldots,\,255\}$. 

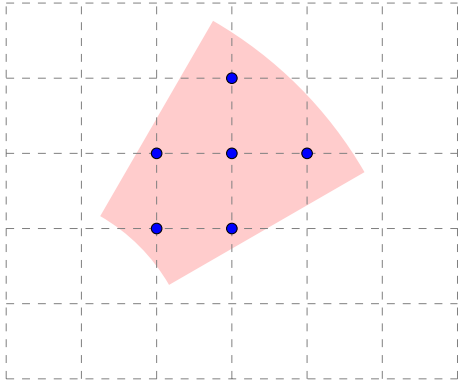
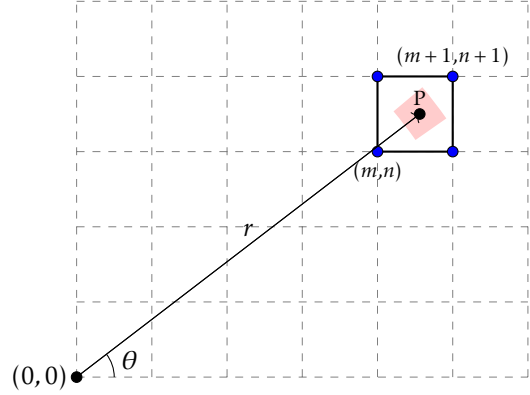
\begin{figure}[hpt]
  \begin{subfigure}{0.45\textwidth}
      \begin{tikzpicture}
\pgfmathsetmacro{\rin}{2.5}
\pgfmathsetmacro{\rout}{5.5}
\pgfmathsetmacro{\thetain}{30}
\pgfmathsetmacro{\thetaout}{60}
\fill[red!20] (\thetain:\rin) arc (\thetain:\thetaout:\rin) -- (\thetaout:\rout) arc (\thetaout:\thetain:\rout);

\draw[step=1cm,gray,ultra thin,dashed] (0,0) grid (6,5);
\draw[fill=blue] (2,2) circle (2pt);
\draw[fill=blue] (3,2) circle (2pt);
\draw[fill=blue] (3,3) circle (2pt);
\draw[fill=blue] (2,3) circle (2pt);
\draw[fill=blue] (4,3) circle (2pt);
\draw[fill=blue] (3,4) circle (2pt);
\end{tikzpicture}
\caption{Computation of pixel intensities based on \meqref{eq_polar_Idef} for $d > 0$, i.e. when sector $S(i,j)$ contains one or more pixels.}
\label{fig:many_pixels}
  \end{subfigure}
\hspace{1cm}
  \begin{subfigure}{0.45\textwidth}
    \begin{tikzpicture}

\draw[step=1cm,gray,ultra thin,dashed] (0,0) grid (6,5);

\draw [thick] (4,3) rectangle (5,4);

\node at (4,3) [below] {\footnotesize{($m$,$n$)}};
\node at (5,4) [above] {\footnotesize{($m+1$,$n+1$)}};
\node at (0,0) [left] {$(0,0)$};

\draw[fill=blue] (4,3) circle (2pt);
\draw[fill=blue] (4,4) circle (2pt);
\draw[fill=blue] (5,3) circle (2pt);
\draw[fill=blue] (5,4) circle (2pt);

\pgfmathsetmacro{\rin}{5.5}
\pgfmathsetmacro{\rout}{6}
\pgfmathsetmacro{\thetain}{35}
\pgfmathsetmacro{\thetaout}{40}
				
\fill[red!20] (\thetain:\rin) arc (\thetain:\thetaout:\rin) -- (\thetaout:\rout) arc (\thetaout:\thetain:\rout);

 \draw[fill=black] (37.5:5.75cm) circle (2pt);
\draw[->] (0,0) -- (37.5:5.75cm) node[right, above] {\footnotesize{P}};

\draw[fill=black] (0,0) circle (2pt);
\draw[fill=black] (0,0) -- (37.5:5.75cm) node[midway, above] {$r$};
\draw (0.5,0) arc (0:37.5:0.5) node[midway, right, yshift=2pt] {$\theta$};

\end{tikzpicture}
\caption{Computation of pixel intensities based on \meqref{eq_polar_Idef} for $d = 0$, i.e. when sector $S(i,j)$ contains no pixels.}
     \label{fig:zero_pixels}
  \end{subfigure}
\caption{Computation of pixel intensities for polar representation of images via smoothing or interpolation (refer \meqref{eq_polar_Idef}).}
\label{fig:sector_pixels}
\end{figure}

We translate our coordinate system such that the origin of the coordinate system is at the center of the image. This means that
 the new coordinates $(i^\prime,j^\prime) $ are given by 
$(i^\prime,j^\prime) = (i - h_1,j - h_2)$,
where for $k=1,\, 2$, 
\begin{equation*}
h_k =
    \begin{cases}
\frac{1}{2} \, \widetilde{N}_k,  & \text{if } \widetilde{N}_k \text{ is even,} \\
\frac{1}{2} \, \left(\widetilde{N}_k -1\right), & \text{if } \widetilde{N}_k \text{ is odd},
\end{cases}
\end{equation*}
or alternatively, $h_k = \floor{\frac{1}{2} \, \widetilde{N}_k}$.
Let $D_x^{\prime}$ and $D_y^{\prime}$ be the sets obtained upon application of this coordinate translation on the sets $D_x$ and $D_y$, respectively. 
Let us fix the maximum radius $r_{max}$ for the polar representation of the image $\widetilde{I}$. It is easy to see that $r_{max} \leq \frac{1}{2}
\min (\widetilde{N}_1, \widetilde{N}_2) $. 
If the input image  $\widetilde{I}$ is not circular, then the pixels in the region outside the circular disk of radius $r_{max}$ are ignored. 
The polar representation of the image $\widetilde{I}$ as an $N_1 \times N_2$ matrix, say $I$, can be defined as follows. 
Let $d$ be the number of pixels of the image $\widetilde{I}$ contained in $S(i,j)$, i.e., $d$ is the cardinality of the set $\mathcal{S}_D := \{(m,n) \in 
\left(D_x^{\prime} \times  D_y^{\prime}\right) \cap S(i,j)\}$.
We set the $(i,j)$th element of the matrix $I$ as 
 \begin{equation} \label{eq_polar_Idef}
    I(i,j) :=  
    \begin{cases}
  &\frac{1}{d} \sum_{(m,n) \in \mathcal{S}_D} \widetilde{I}(m,n), \quad   \text{ if } d > 0, \\
 & \text{Average of the pixel intensities of pixels lying in }  B_{1}(m,n), \quad \text{ if } d = 0,\\
\end{cases}
\end{equation}
where $B_{1}(m,n) \subset \left(D_x^{\prime} \times  D_y^{\prime}\right)$ is an open disc of radius $1$ centered at $(m,n)$. Here, $(m,n)$
denotes the Cartesian coordinate of the centroid of the annular region  $S(i,j) $.
We note that if the annular region defined by $S(i,j)$ contains one or more pixels of the input image $\widetilde{I}$, then 
$I(i,j)$ is assigned the average of the intensities of all the pixels contained in the annular region defined by $S(i,j)$ (as shown in \mfig{fig:many_pixels}).
If the annular region $S(i,j)$ does not contain any pixels of the input image $\widetilde{I}$, then $I(i,j)$ is computed as the average of the intensities of the nearest neighboring pixels relative to the centroid of the annular region defined by $S(i,j)$ (as shown in \mfig{fig:zero_pixels}). Therefore, for the case in \mfig{fig:zero_pixels} we have
\begin{equation*}
    I(i,j) = \frac{1}{4}\left(\widetilde{I}(m,n) + \widetilde{I}(m+1,n) + \widetilde{I}(m,n+1) + \widetilde{I}(m+1,n+1)\right).
\end{equation*}

It is clear from \meqref{eq_polar_Idef} that qualitatively, the polar representation of a Cartesian image is like applying a smoothening filter as it involves averaging (or smoothening) of pixel intensities over the sectors $S(i,j)$. Consider an input image $\widetilde{I}$
of size $\widetilde{N}_1 \times \widetilde{N}_2 = 512 \times 512$. If we represent it as polar image of size $N_1 \times N_2  =  2\times 2$, given by the matrix 
\[
\mat{I(0,0)}{I(0,1)}{I(1,0)}{I(1,1)},
\]
then $I(0,0)$, $I(0,1)$, $I(1,0)$ and $I(1,1)$ represent the average intensities of the pixels lying inside the annular regions $S(0,0)$, $S(0,1)$, $S(1,0)$ and $S(1,1)$, respectively (refer to \mfig{polar_discretization}). In case one is interested in ensuring that the map from the Cartesian representation to the polar representation of the image (ref. \meqref{eq_polar_Idef}) based on \uarea \, is approximately one-to-one, then the following condition must hold: $$
\widetilde{N}_1 \times \widetilde{N}_2 \approx \pi r_{max}^2 \approx N_1 \times N_2.  
$$ 
As noted earlier, the \uarea \, ensures that each of the sectors $S(i, j)$ contains approximately the same number of pixels.

Based on the discussion above, an algorithm to obtain the polar representation of an image from its Cartesian representation is presented in Algorithm \ref{alg_image_polar_representation}.

\begin{algorithm}[H] \label{alg_image_polar_representation}
    \DontPrintSemicolon
    \KwInput{ \\
        (a) A Cartesian representation of an image as a matrix  $\widetilde{I}$ of size $\widetilde{N}_1\times\widetilde{N}_2$. Here  $1 ~{<}~ \widetilde{N}_1 \in \NN$, $1 ~{<}~ \widetilde{N}_2\in \NN$. \\
        (b) Parameter $f$, where $f=\frac{1}{2}$ corresponds to \uarea \, and $f=1$ corresponds to \uradial. \\
        (c) $N_1$ and $N_2$, where $N_1 \times N_2$ is the desired size of the matrix for the corresponding polar representation of the image. Here $1~ {<} ~ N_1 \in \NN$, $1 ~{<}~ N_2 \in \NN$. Refer \mref{subsec:polar_representation}
}
	\KwOutput{
 An $N_1\times N_2$ matrix $I$ corresponding to the polar representation of the input image $\widetilde{I}$.  }
\Fn{ CartesianToPolar($\widetilde{I}$, $N_1$, $N_2$, $f$)} {
Initialize two matrices, namely `\textit{sum}' and `\textit{count}', both of size $N_1 \times N_2$, with all their elements set to $0$.\\
$r_{max} = \frac{1}{2}\mbox{min}(\widetilde{N}_1,\,\widetilde{N}_2)$ \tcp*{the pixels in the region outside the open circular disk of radius $r_{max}$ are ignored.}
	\For{$ i \gets 0$ \KwTo $\widetilde{N}_2 -1$ }
	{
		\For{$ j \gets 0$ \KwTo $\widetilde{N}_2 -1$ }
		{
		    $ i' = i- \floor{\frac{\widetilde{N}_1}{2}} $, \hspace{0.5mm} $ j' = j- \floor{\frac{\widetilde{N}_2}{2}} $ \tcp*{Shifts $i$, $j$ to centre of the image.}
      Compute the polar coordinates $r \in [0, r_{max})$ and $\theta  \in   [0, 2 \pi)$ corresponding to the Cartesian coordinates $(i',j')$. 
      \tcp*{Calculates radii and angle for individual matrix element.}
		    $k$ = $\left\lfloor \left(\frac{r}{r_{max}}\right)^{1/f} \cdot N_1\right\rfloor$ , \hspace{0.5mm} $q$ = $\left\lfloor \frac{\theta}{2\pi} \cdot N_2\right\rfloor$ 
       \tcp*{The pixel with Cartesian coordinates $(i,j)$ lies in the annular region or sector $S(k,q)$.}
		    \If{$k<N_1$ \textbf{and} $q<N_2$}
		    {   
                \textit{sum}($k$, $q$) = \textit{sum}($k$, $q$) + $\widetilde{I}(i, j)$ \\   
	            \textit{count}($k$, $q$) = \textit{count}($k$, $q$) + 1
	        }
	    }
	}
    
	\For{$ k \gets 0$ \KwTo $N_1 -1$ }
	{
		\For{$ q \gets 0$ \KwTo $ N_2- 1$ }
		{
        \If{\textit{count}($k$, $q$) $\neq$ 0}
            {
    	       $I(k, q)$ = $ \left\lfloor \textit{sum}(k, \, q) / \textit{count}(k, \, q) \right\rfloor$ \tcp*{Each element of the array \textit{sum} is divided by its corresponding element in the array \textit{count}.}  
            }
            {
               $r = \left\lfloor\frac{r_{max}}{2}\left(\left(\frac{k}{N_1}\right)^f + \left(\frac{1+k}{N_1}\right)^f\right)\right\rfloor$,\hspace{0.5mm} $\theta = \left\lfloor\frac{\pi}{N_2}(1+2q)\right\rfloor$\\
                $m =  r \, \cos \theta + \frac{\widetilde{N}_1}{2},~n =  r \,  \sin \theta + \frac{\widetilde{N}_2}{2} $\\
                $I$($k$, $q$) = Average of the pixel intensities of all the pixels in $B_{1}(m,n)$, which is the open disk of radius $1$ centered at $(m,n)$.} \tcp*{Refer to \meqref{eq_polar_Idef}}
               
            }
        }	
 \Return{$I$.}
 }
 \caption{An algorithm to obtain the polar representation of an image from its Cartesian representation, using the framework discussed in \mref{subsec:polar_representation}.}
\end{algorithm}

\vspace{0.2cm}

\subsection{Conversion from polar image representation to its Cartesian form} \label{sec:conversion_pc}

 %
In this section, we provide algorithm for the conversion of an image from its polar representation to its Cartesian representation by employing the framework described in \mref{subsec:polar_representation}. 
Maintaining the notation unchanged, let the input image matrix $I$ (of size $N_1 \times N_2$) denote the image in polar representation, and the corresponding output image in Cartesian coordinates is denoted by matrix $\widetilde{\widetilde{I}}$ of size $\widetilde{N}_1\times \widetilde{N}_2$. It is worth noting that, during the conversion of the image from Cartesian representation to polar representation, the parameter $r_{max}$ plays a crucial role (refer to \meqref{area_and_radial_distribution} and \meqref{Walsh_radial_distribution}). Only the pixels inside an open disk $D$ of radius $r_{max}$ are considered during this conversion process. Therefore, the data in polar representation depends on the choice of $r_{max}$. When converting back to Cartesian coordinate representation from polar representation, the same $r_{max}$ must be used. Bearing this in mind, it is observed that the matrix elements of $\widetilde{\widetilde{I}}$ are given by:

\begin{equation}\label{eq_cart_Idef}
    \widetilde{\widetilde{I}}(i,j) =  I(m,n), 
\end{equation}

where $m\in \mathbb{M}_x$, $n\in \mathbb{M}_y$, and $(i,j)\in S(m,n)$ (refer to \meqref{eq:skq}).
Algorithm \ref{alg_image_Cartesian_representation} captures the procedure described above for transforming the image from its polar representation to its Cartesian representation.

\begin{algorithm}[H] \label{alg_image_Cartesian_representation}
    \DontPrintSemicolon
    \KwInput{\\
    (a) A polar representation of an image as a matrix $I$ of size $N_1\times N_2$. Here $1 ~{<}~ N_1 \in \NN$, $1 ~{<}~ N_2 \in \NN$. \\
    (b) Parameter $f$, where $f=\frac{1}{2}$ corresponds to \uarea \, and $f=1$ corresponds to \uradial. \\
    (c) $ \widetilde{N}_1$ and $\widetilde{N}_2$, where $ \widetilde{N}_1$ and $\widetilde{N}_2$ are the row and column sizes of the image in the Cartesian representation.  Here  $1 \leq \widetilde{N}_1 \in \NN$, $1 \leq \widetilde{N}_2\in \NN$.
    }
	\KwOutput{Matrix $\widetilde{\widetilde{I}}$ of size $\widetilde{N}_1\times \widetilde{N}_2$ corresponding to the Cartesian representation of the input image $I$.}
  \Fn{\textbf{PolarToCartesian} ($I$, $\widetilde{N}_1$, ~$\widetilde{N}_2$, $f$)}{
    Initialize a matrix of size $\widetilde{N}_1 \times \widetilde{N}_2$ with each matrix element set to 255.  \tcp*{On a grayscale 255 corresponds to white color.}
    $r_{max} = \left\lfloor\frac{1}{2}\mbox{min}(\widetilde{N}_1,\,\widetilde{N}_2)\right\rfloor$ \tcp*{the pixels in the region outside the circular disk of radius $r_{max}$ are ignored.}
	\For{$ i \gets 0$ \KwTo $ \widetilde{N}_1-1$ }
	{
		\For{$ j \gets 0$ \KwTo $ \widetilde{N}_2-1$ }
		{
       $ i' = i- \floor{\frac{\widetilde{N}_1}{2}} $, \hspace{0.5mm} $ j' = j- \floor{\frac{\widetilde{N}_2}{2}} $ \tcp*{Shifts $i$, $j$ to centre of the image.}
      Compute the polar coordinates $r \in [0, r_{max})$ and $\theta  \in   [0, 2 \pi)$ corresponding to the Cartesian coordinates $(i',j')$. 
      \tcp*{Calculates radii and angle for individual matrix element.}
            $k$ = $\left\lfloor \left(\frac{r}{r_{max}}\right)^{1/f} \cdot N_1\right\rfloor$ , \hspace{0.5mm} $q$ = $\left\lfloor \frac{\theta}{2\pi} \cdot N_2\right\rfloor$ \tcp*{$\left\lfloor \cdot \right\rfloor$ symbolizes floor function.}
		    \If{$k<N_1$ \textbf{and} $q<N_2$}
		    {
		        $\widetilde{\widetilde{I}}(i,j)$ = $I(k,q)$ \\   
	        }
	    }
	}  
	\Return{$\widetilde{\widetilde{I}}$.} 
 }
	\caption{An algorithm for obtaining the image in Cartesian representation from its polar representation using \meqref{eq_cart_Idef}.} 
\end{algorithm}

\begin{figure}[ht]
	\begin{center}
     \begin{subfigure}[t]{.3\textwidth}
			\centering
			\includegraphics[width=\linewidth]{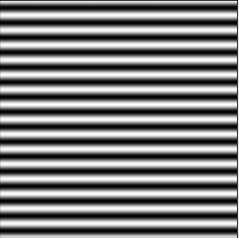}
		    \caption{Horizontal uniform bands. }\label{horizontal_banding_noise}
		\end{subfigure}
  \hspace{0.5cm}
        \begin{subfigure}[t]{.3\textwidth}
			\centering
			\includegraphics[width=\linewidth]{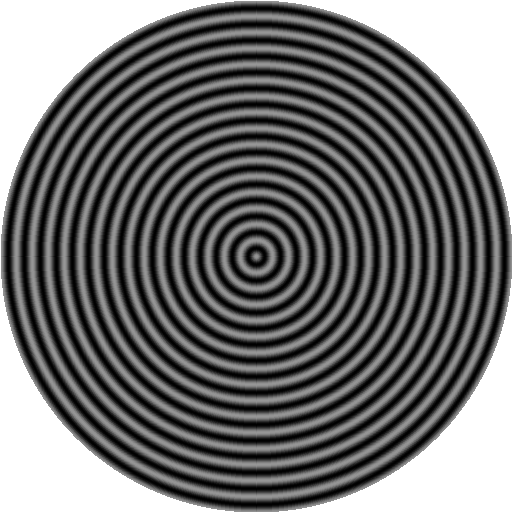}
			\caption{Circular bands with \uradial.}\label{img_uniform_radial_noise}
		\end{subfigure}
        \hspace{0.5cm}
        \begin{subfigure}[t]{.3\textwidth}
			\centering
			\includegraphics[width=\linewidth]{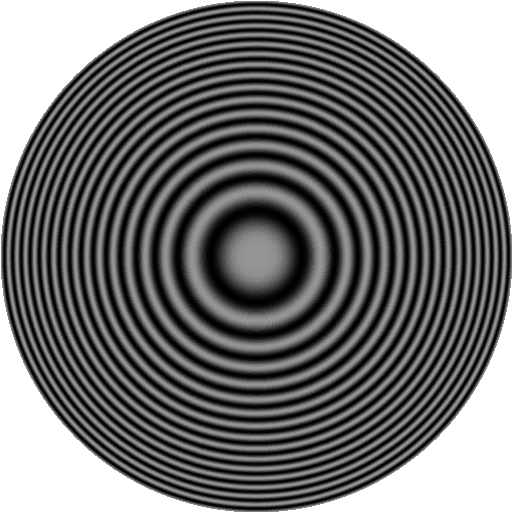}
		    \caption{Circular bands with \uarea.}\label{img_WH_radial_noise}
		\end{subfigure}
        \hspace{0.5cm}
       \\ \vspace{0.5cm}
      \begin{subfigure}[t]{.3\textwidth}
			\centering
			\includegraphics[width=\linewidth]{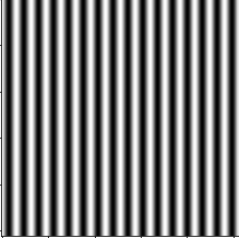}
			\caption{Vertical uniform bands.}\label{vertical_banding_noise}
		\end{subfigure}
      \hspace{0.5cm}
   \begin{subfigure}[t]{.3\textwidth}
			\centering
			\includegraphics[width=\linewidth]{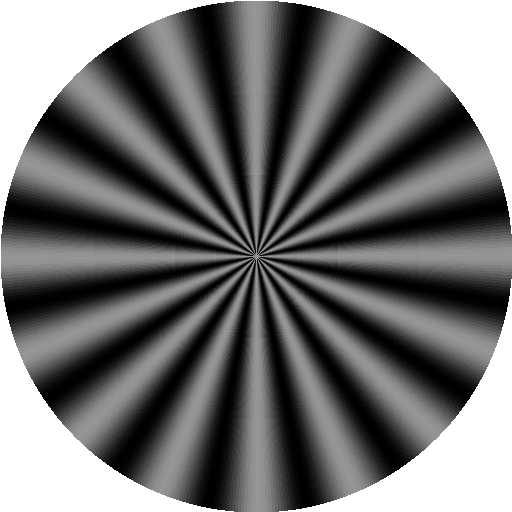}
		    \caption{Uniform azimuthal bands.}\label{img_theta_banding_noise}
		\end{subfigure}
\end{center}
\caption{Images with vertical and horizontal bands are shown on the left. The images on the right are polar representations of the images on the left.} \label{Cartesian_to_polar}
\end{figure}

Computational examples illustrating conversion from Cartesian representations to polar representations of images are provided in \mfig{Cartesian_to_polar}. The images on the left in \mfig{Cartesian_to_polar} contain horizontal and vertical bands (shown in \mfig{horizontal_banding_noise} and \mfig{vertical_banding_noise}, respectively). The application of Algorithm \ref{alg_image_polar_representation}, with $f=1$ (i.e., using the \uradial), on \mfig{horizontal_banding_noise} results in an image with periodic bands along radial directions and is shown in \mfig{img_uniform_radial_noise}.  
The image shown in \mfig{img_WH_radial_noise} results from an application of Algorithm \ref{alg_image_polar_representation} with $f=\frac{1}2$ (i.e., using the \uarea) on the image shown in \mfig{horizontal_banding_noise} (containing the horizontal bands). Henceforth, we will refer to periodic bands along the radial direction (\mfig{img_uniform_radial_noise} and \mfig{img_WH_radial_noise}) as circular bands. Application of Algorithm \ref{alg_image_polar_representation} with $f = \frac{1}{2}$ on vertical bands (shown in \mfig{vertical_banding_noise}) results in azimuthal bands shown in \mfig{img_theta_banding_noise}. In the other direction, Algorithm
\ref{alg_image_Cartesian_representation} can be used (with an appropriate choice of the parameter $f$) to convert the images containing circular and azimuthal bands to their Cartesian representations resulting in images containing vertical and horizontal bands, respectively. 

\section{Hybrid classical-quantum algorithm for removal of periodic banding noises}\label{sec:removal_circular_noise}
Periodic banding noise in images manifests as regular alternating dark and light stripes or bands, often resulting from interference or errors during image acquisition or processing. These disruptive bands can impact the overall visual quality of the image. 
\mfig{Cartesian_to_polar} displays images with vertical, horizontal, circular, and azimuthal bands. When these images mix with a given image of interest, they create periodic banding noise. 
Of course, banding noises in images are often undesirable. 

\begin{figure}[ht]
	\begin{center}
    \includegraphics[scale=0.5]{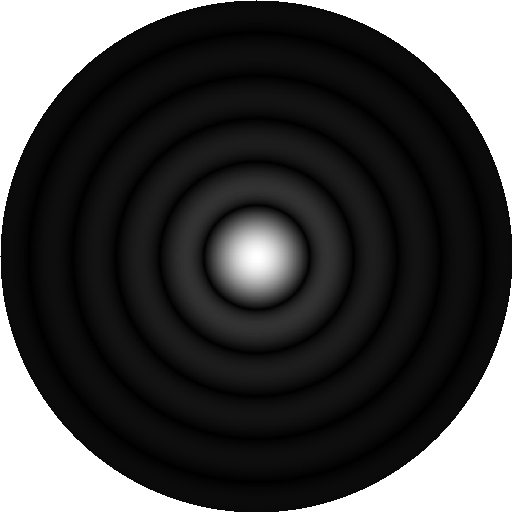}
    \caption{A computer-generated grayscale image for Airy Pattern (using $ka = 2 \pi$).  Intensities of the grayscale image have been adjusted such that the brightness of the outer rings of the Airy pattern is enhanced.}\label{img:img_airy_pattern}
  \end{center}
\end{figure}

Another important pattern of circular banding noise is the Airy pattern \cite{airy1835diffraction,born2013principles}. An airy pattern is formed when a circular aperture is uniformly illuminated with light whose wavelength is comparable to the radius of a circular aperture. The phenomenon that produces airy pattern is caused by Fraunhofer diffraction. The intensity distribution of concentric discs in the Airy pattern is 
\begin{equation}\label{eq_airy_pattern_intensity}
    I(\theta) = I_0\left(\frac{2J_1(k \, a \, \sin\theta)}{k \, a \, \sin\theta}\right)^2.
\end{equation}
Here $J_1(x)$ is Bessel's function of the first kind of order one, $k$ describes the wave number associated with the light used to illuminate the circular aperture, and $a$ is the radius of the circular aperture. $\theta$ is the angle between the axis passing through the center of the circular aperture and the line between aperture center and observation point on the screen.  \mfig{img:img_airy_pattern} shows computationally generated grayscale of the airy pattern from \meqref{eq_airy_pattern_intensity}. 

\vspace{0.5cm}

In this work, we propose a hybrid classical-quantum approach for removal of circular (including Airy pattern) and azimuthal banding noises using the polar representation of the images as described earlier in 
\mref{subsec:polar_representation}.

\begin{table}[htb]
    \centering
    \begin{tabular}{cccc}
        \toprule
        Banding Noise Type &  Period &  Noise Matrix $M$ & $\QWHT(M)$  \\
        \midrule
        Circular  & $2$ & $ \begin{bmatrix}
        z&z&z&z&z&z&z&z\\
        0&0&0&0&0&0&0&0\\
        z&z&z&z&z&z&z&z\\
        0&0&0&0&0&0&0&0\\
        z&z&z&z&z&z&z&z\\
        0&0&0&0&0&0&0&0\\
        z&z&z&z&z&z&z&z\\
        0&0&0&0&0&0&0&0\\
    \end{bmatrix} $  & $ 4\begin{bmatrix}
        z&0&0&0&0&0&0&0\\
        0&0&0&0&0&0&0&0\\
        0&0&0&0&0&0&0&0\\
        0&0&0&0&0&0&0&0\\
        0&0&0&0&0&0&0&0\\
        0&0&0&0&0&0&0&0\\
        0&0&0&0&0&0&0&0\\
        z&0&0&0&0&0&0&0
    \end{bmatrix} $ \\
    \hline \\
     Circular  & $4$ &
       $\begin{bmatrix}
 	z  &       z  &       z  &       z  &       z  &       z  &       z  &       z \\
 	z  &       z  &       z  &       z  &       z  &       z  &       z  &       z \\
 	0  &       0  &       0  &       0  &       0  &       0  &       0  &       0 \\
 	0  &       0  &       0  &       0  &       0  &       0  &       0  &       0 \\
 	z  &       z  &       z  &       z  &       z  &       z  &       z  &       z \\
 	z  &       z  &       z  &       z  &       z  &       z  &       z  &       z \\
 	0  &       0  &       0  &       0  &       0  &       0  &       0  &       0 \\
 	0  &       0  &       0  &       0  &       0  &       0  &       0  &       0 
 \end{bmatrix}$ &
 $4\begin{bmatrix}
 	z  &       0  &       0  &       0  &       0  &       0  &       0  &       0 \\
 	0  &       0  &       0  &       0  &       0  &       0  &       0  &       0 \\
 	0  &       0  &       0  &       0  &       0  &       0  &       0  &       0 \\
 	z  &       0  &       0  &       0  &       0  &       0  &       0  &       0 \\
 	0  &       0  &       0  &       0  &       0  &       0  &       0  &       0 \\
 	0  &       0  &       0  &       0  &       0  &       0  &       0  &       0 \\
 	0  &       0  &       0  &       0  &       0  &       0  &       0  &       0 \\
 	0  &       0  &       0  &       0  &       0  &       0  &       0  &       0 
 \end{bmatrix} $\\
        \hline \\
  Azimuthal & 2 &  $ \begin{bmatrix}
	z  &       0  &       z  &       0  &       z  &       0  &       z  &       0 \\
	z  &       0  &       z  &       0  &       z  &       0  &       z  &       0 \\
	z  &       0  &       z  &       0  &       z  &       0  &       z  &       0 \\
	z  &       0  &       z  &       0  &       z  &       0  &       z  &       0 \\
	z  &       0  &       z  &       0  &       z  &       0  &       z  &       0 \\
	z  &       0  &       z  &       0  &       z  &       0  &       z  &       0 \\
	z  &       0  &       z  &       0  &       z  &       0  &       z  &       0 \\
	z  &       0  &       z  &       0  &       z  &       0  &       z  &       0 
\end{bmatrix}$  &    4 $ \begin{bmatrix}
	z  &       0  &       0  &       0  &       0  &       0  &       0  &       z \\
	0  &       0  &       0  &       0  &       0  &       0  &       0  &       0 \\
	0  &       0  &       0  &       0  &       0  &       0  &       0  &       0 \\
	0  &       0  &       0  &       0  &       0  &       0  &       0  &       0 \\
	0  &       0  &       0  &       0  &       0  &       0  &       0  &       0 \\
	0  &       0  &       0  &       0  &       0  &       0  &       0  &       0 \\
	0  &       0  &       0  &       0  &       0  &       0  &       0  &       0 \\
	0  &       0  &       0  &       0  &       0  &       0  &       0  &       0 
\end{bmatrix}$\\
\hline \\
    Azimuthal & 4 & $ \begin{bmatrix}
	z  &       z  &       0  &       0  &       z  &       z  &       0  &       0 \\
	z  &       z  &       0  &       0  &       z  &       z  &       0  &       0 \\
	z  &       z  &       0  &       0  &       z  &       z  &       0  &       0 \\
	z  &       z  &       0  &       0  &       z  &       z  &       0  &       0 \\
	z  &       z  &       0  &       0  &       z  &       z  &       0  &       0 \\
	z  &       z  &       0  &       0  &       z  &       z  &       0  &       0 \\
	z  &       z  &       0  &       0  &       z  &       z  &       0  &       0 \\
	z  &       z  &       0  &       0  &       z  &       z  &       0  &       0 
\end{bmatrix}$
&
$4\begin{bmatrix}
	z  &       0  &       0  &       z  &       0  &       0  &       0  &       0 \\
	0  &       0  &       0  &       0  &       0  &       0  &       0  &       0 \\
	0  &       0  &       0  &       0  &       0  &       0  &       0  &       0 \\
	0  &       0  &       0  &       0  &       0  &       0  &       0  &       0 \\
	0  &       0  &       0  &       0  &       0  &       0  &       0  &       0 \\
	0  &       0  &       0  &       0  &       0  &       0  &       0  &       0 \\
	0  &       0  &       0  &       0  &       0  &       0  &       0  &       0 \\
	0  &       0  &       0  &       0  &       0  &       0  &       0  &       0 
\end{bmatrix}$ \\
        \bottomrule
    \end{tabular}
    \caption{The table illustrates characteristic patterns in the transformed domain (as shown in the last column) for circular and azimuthal banding noises in polar representation of different periods. The transformed matrices $\QWHT(Z)$ contain non-zero entries in the first column (for circular banding noise) and first row (for azimuthal banding noise), respectively, with all other elements being $0$.}
    \label{tab:example}
\end{table}

To see how the algorithm works, consider a $N\times N$ ($N=2^n,n\in \mathbb{N}$) matrix $M$ representing a noisy image containing the periodic circular and/or azimuthal banding noises. The goal is to separate out the parts representing the periodic noise $Z$ and the noise-free image $A$. Since the noises considered here are additive, we have $M = A + Z$,
where the noise matrix $Z$ of size $N\times N$ ($N=2^n,n\in \mathbb{N}$) is either a matrix containing circular or azimuthal banding noise. Let $Z_c$ denote the circular banding noise and $Z_a$ denote the azimuthal banding noise. We note that a circular banding noise of period $2$ may manifest itself in a matrix $Z_c$ of the form 
\begin{equation*}
        Z_c(i,j) = 
        \begin{cases}  z, \quad \text{if } i\mod 2 \equiv 0, \, 0\leq j \leq N-1, \\ 
                      0, \quad \text{if } i\mod 2 \equiv 1, \, 0\leq j \leq N-1,
    \end{cases}
\end{equation*}
where all the entries in the odd rows are $0$ and all the entries in the even rows are $z$.
The application of the Walsh-Hadamard transform, using Algorithm  \ref{alg_QWHT}, results in a matrix  $\QWHT(Z_h)$,  such that 
\begin{equation*}
      \QWHT(Z_h) = 
        \begin{cases}  \frac{Nz}{2},  \quad &\text{if }  (i,j) = (0,0)  \text{ or }   (i,j) = (N-1,0),  \\ 
                    0, \quad &\text{otherwise}.
    \end{cases}
\end{equation*}
It means all the elements of the transformed matrix  $\QWHT(Z_h)$ are $0$, except the first and the last elements of the first column. It can be checked that for circular banding noises of different periods, the non-zero entries in the corresponding transformed matrix  $\QWHT(Z_h)$ are always concentrated in the first column. Similarly, one can verify that for 
azimuthal banding noises of different periods, the non-zero entries in the corresponding transformed matrix  $\QWHT(Z_a)$ are always concentrated in the first row. Some such examples are provided in Table \ref{tab:example}. 
The above observation provides an approach for suppressing the circular and azimuthal banding noise present in an image.  Application of the hybrid classical-quantum Walsh-Hadamard transform (as given in Algorithm \ref{alg_two_dimensional_Walsh_Transfrom}) on the polar representation of the image with circular or azimuthal banding noise, followed by suppressing all the elements in the first row or first column, respectively, except for the $(0,0)$-th element of the matrix, and finally upon performing an inverse Walsh-Hadamard transform,  reduces the noise in the image. A similar method can be used for the removal of the azimuthal banding noise. The approach described above is captured and presented as an algorithm in Algorithm \ref{alg_circular_noise_removal}.

\begin{algorithm}[H] \label{alg_circular_noise_removal}
    \DontPrintSemicolon
     \KwInput{ \\
    (a) A circular noisy grayscale image $\widetilde{I}$, of radius $r_{max}$, given in Cartesian coordinates. If the input image is not circular and is of size $\widetilde{N}_1 \times \widetilde{N}_2 $, then the pixels in the region outside the circular disk $r_{max} = \left\lfloor \frac{1}{2} \, \text{min} (\widetilde{N}_1,\widetilde{N}_2)\right\rfloor$ are ignored.  
    \\
    (b) $N_1$ and $N_2$, representing the number of radial and angular regions or sectors in the polar representations used as an intermediate step in the algorithm. Note that $N_1$ and $N_2$ must be of the form $N_1 = 2^{n_1}$ and $N_2 = 2^{n_2}$, with $1 \leq n_1 \in \NN$, $1 \leq n_2 \in \NN$.\\
    (c) Parameter $f$, where $f=\frac{1}{2}$ corresponds to \uarea \, and $f=1$ corresponds to \uradial. \\
    (d) Input flag, \textit{cflag},  where \textit{cflag} is set to $1$ if the removal of circular banding noise from the noisy input image $\widetilde{I}$ is desired,  otherwise it is set to $0$.  \\
    (e) Input flag, \textit{aflag}, where \textit{aflag} is set to $1$ if the removal of azimuthal banding noise from the noisy input image $\widetilde{I}$ is desired, otherwise it is set to $0$. 
    }
	\KwOutput{Grayscale image $I$ of radius $r_{max}$ given in Cartesian coordinates with circular and/or azimuthal banding noise filtered.}
        Set $\widetilde{N}_1$ and $\widetilde{N}_2$ as the row and column sizes, respectively, of the image matrix $\widetilde{I}$. \\
	{$X$ = \textbf{CartesianToPolar}($\widetilde{I}$,~$N_1$,~$N_2$, $f$)}   \tcp*{Converting image to discretized polar coordinates using Algorithm \ref{alg_image_polar_representation}.}
	{ $ X = \Qwalsh^{\otimes 2} $($ X $)}   \tcp*{Compute the two-dimensional Walsh-Hadamard transform of $ X $ using Algorithm~\ref{alg_two_dimensional_Walsh_Transfrom}.}
\If{ \textit{cflag} is $1$} 
{
 \For{$ i \gets 1$ \KwTo $ N_1-1$ }
	{
		$  X(i,0) = 0 $	  \tcp*{Suppress all but the first element in the first column in the sequency domain}
	}
 }
 \If{ \textit{aflag} is $1$} 
 {
 \For{$ j \gets 1$ \KwTo $ N_2-1$ }
	{
		$ X(0,j) = 0$	  \tcp*{Suppress all but the first element in the first column in the sequency domain}
	}
 } 
	{ $ X = \Qwalsh^{\otimes 2} $($ X $)} \tcp*{Compute the two-dimensional inverse Walsh-Hadamard transform of $ X $ using Algorithm~\ref{alg_two_dimensional_Walsh_Transfrom}.}
    {$I$ = \textbf{PolarToCartesian}($X$,~$\widetilde{N}_1$,~$\widetilde{N}_2$, $f$)}   \tcp*{Converting image to Cartesian coordinates using Algorithm \ref{alg_image_Cartesian_representation}.}
	\Return{$ I $.}
	\caption{An algorithm for removal of circular banding noise}
\end{algorithm}

\begin{remark}
	\leavevmode
		\begin{enumerate}[i.] 
            \item While converting between polar and Cartesian representations, the sizes of matrix rows and columns are not required to be powers of $2$. However, for the quantum Walsh-Hadamard transform, both the number of rows and columns in the matrix must be powers of $2$.
            \item The input flag `\textit{cflag}' denotes the presence of circular banding noise, with `\textit{cflag}' set to $1$ if the noise is present, and to $0$ otherwise, while the input flag `\textit{aflag}' indicates the presence of azimuthal banding noise, with `\textit{aflag}' set to $1$ if the noise is present, and to $0$ otherwise. For example, for an image containing both types of noise,  both input flags are set to $1$, for removing noise.
		\end{enumerate}
\end{remark}

 \subsection{Computational examples} \label{sec:computational_example}
\begin{figure}[ht]
	\begin{center}
			\includegraphics[scale=0.4]{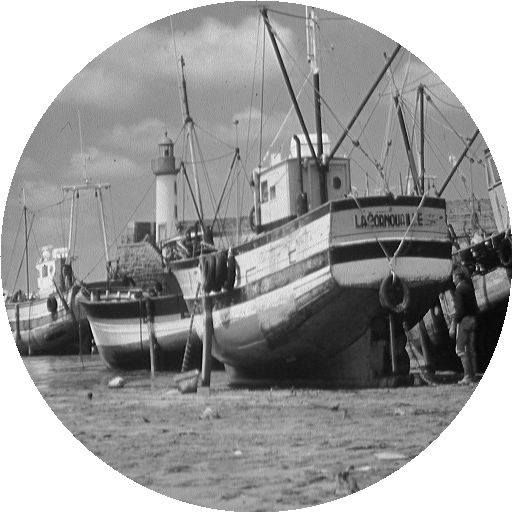}
			\caption{Noise free original image.} \label{fig:img:original}
    \end{center}
\end{figure}

 We present several computational examples to demonstrate the application of Algorithm \ref{alg_circular_noise_removal} in filtering noisy images containing: (a) circular banding noise (for both \uradial \, and \uarea cases), (b) azimuthal banding noise and (c) Airy pattern noise. For these examples, noisy images were created by adding different types of noises to the original image shown in \mfig{fig:img:original}. 
The original and the noisy images considered were of the size $512 \times 512$. All the noisy images were filtered using Algorithm \ref{alg_circular_noise_removal} and after removing the noise   
 the filtered images in Cartesian representation were obtained. 
 While using Algorithm \ref{alg_circular_noise_removal}, the noisy input images of the size $\widetilde{N}_1 \times \widetilde{N}_2 = 512 \times 512$ were converted to their corresponding polar representations of the size $N_1 \times N_2 = 256 \times 512$, i.e., the input values $N_1 = 256$ and $N_2 = 512$ were used in Algorithm \ref{alg_circular_noise_removal}.
 The resulting filtered images of the size $\widetilde{N}_1 \times \widetilde{N}_2 = 512 \times 512$ were compared with the original image, \mfig{fig:img:original},  using standard image quality metrics such as Structural Similarity Index Measure (SSIM), Peak Signal-to-Noise Ratio (PSNR) and Mean Square Error (MSE). More details on these image quality metrics can be found in \cite{shukla2022hybrid}. These examples were implemented and tested using the simulation environment on  IBM's open source quantum computing platform Qiskit.

\subsubsection{Circular banding noise}
\begin{figure}[ht]
	\begin{center}
		\begin{subfigure}[t]{.3\textwidth}
			\centering
			\includegraphics[width=\linewidth]{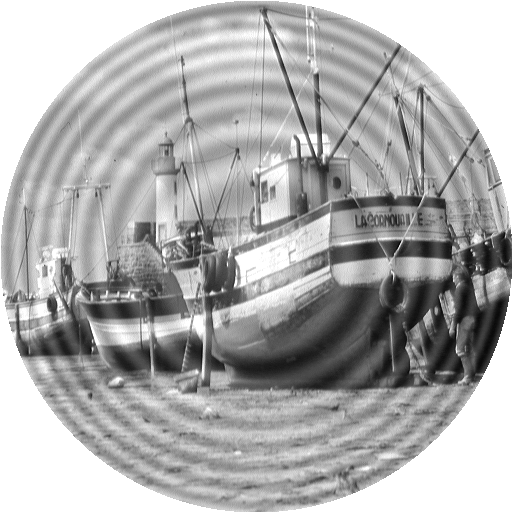}
			\caption{Noisy image.}\label{img:noisy_img_uarea}
		\end{subfigure}~~~~~~~~~~~
        \begin{subfigure}[t]{.3\textwidth}
			\centering
			\includegraphics[width=\linewidth]{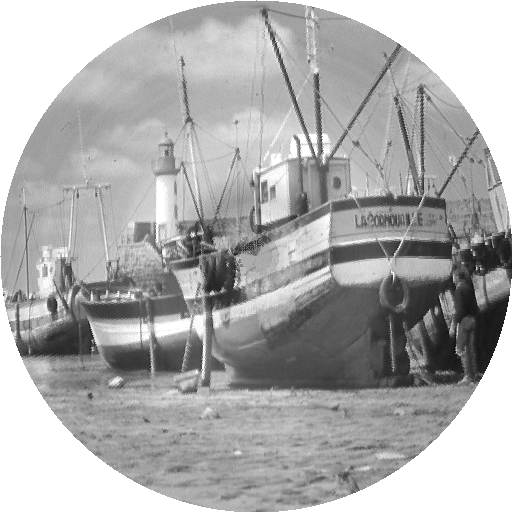}
			\caption{Image filtered with $f$ = 0.5, SSIM = 0.90,  PSNR = 26.13, MSE = 638.55.}\label{img:noisy_img_uarea_filtered_f=0.5}
		\end{subfigure}~~~~~~~~~~~
        \begin{subfigure}[t]{.3\textwidth}
			\centering
			\includegraphics[width=\linewidth]{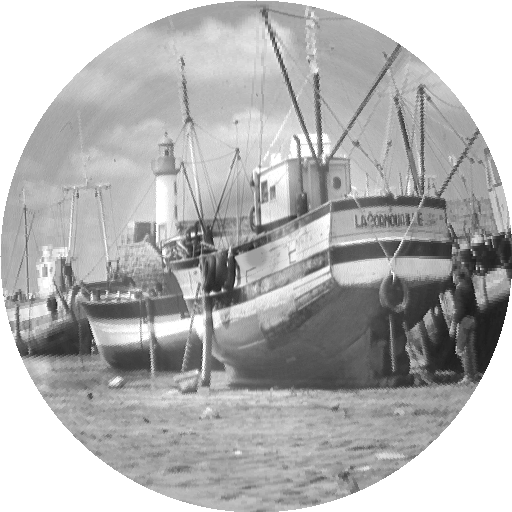}
			\caption{Image filtered with $f$ = 1.0, SSIM =0.88 ,  PSNR = 24.43, MSE = 944.84.}\label{img:noisy_img_uarea_filtered_f=1.0}
		\end{subfigure}
  \caption{Filtering of the circular banding noise (associated with \uarea). } \label{img:noisy_circular_uarea}
    \end{center}
\end{figure}

\begin{figure}[ht]
	\begin{center}
		\begin{subfigure}[t]{.3\textwidth}
			\centering
			\includegraphics[width=\linewidth]{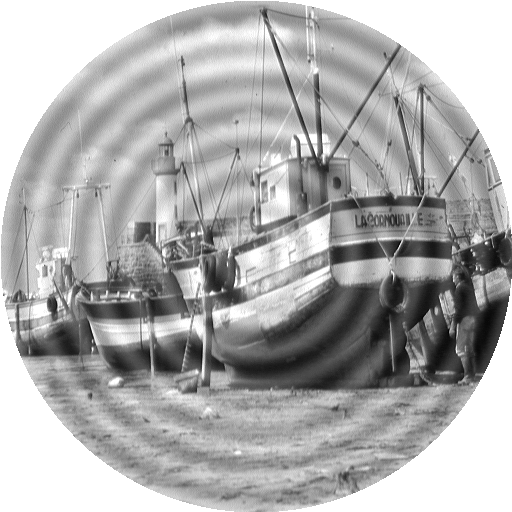}
			\caption{Noisy image.}\label{img:noisy_img_uradial}
		\end{subfigure}~~~~~~~~~~~
        \begin{subfigure}[t]{.3\textwidth}
			\centering
			\includegraphics[width=\linewidth]{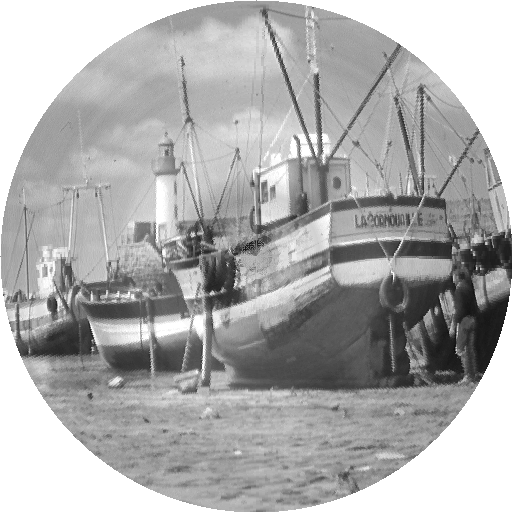}
			\caption{Image filtered with $f$ = 0.5, SSIM = 0.90 ,  PSNR = 26.14, MSE = 637.55.}\label{img:noisy_img_uradial_filtered_f=0.5}
		\end{subfigure}~~~~~~~~~~~
        \begin{subfigure}[t]{.3\textwidth}
			\centering
			\includegraphics[width=\linewidth]{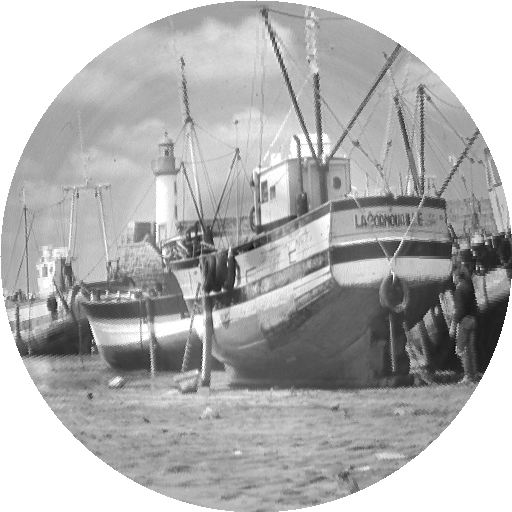}
			\caption{Image filtered with $f$ = 1.0, SSIM = 0.88,  PSNR = 24.84, MSE = 859.66.}\label{img:noisy_img_uradial_filtered_f=1.0}
		\end{subfigure}
  \caption{Filtering of the circular banding noise (associated with \uradial). } \label{img:noisy_circular_uradial}
    \end{center}
\end{figure}

Examples demonstrating the filtering of the images with circular banding noise associated with \uarea \, and  \uradial \, are shown in \mfig{img:noisy_circular_uarea} and \mfig{img:noisy_circular_uradial}, respectively. We note that \mfig{img:noisy_img_uarea} displays a noisy image containing the circular banding noise associated with \uarea. The application of Algorithm \ref{alg_circular_noise_removal}, for the cases of input parameters $f = 0.5$ (corresponding to \uarea) and $f = 1$ (corresponding to \uradial), resulted in filtered images. These filtered images are displayed in \mfig{img:noisy_img_uarea_filtered_f=0.5} and \mfig{img:noisy_img_uarea_filtered_f=1.0} along with image quality metrics such as SSIM, PSNR, and MSE.

Similarly, the application of Algorithm \ref{alg_circular_noise_removal} for filtering of the image \mfig{img:noisy_img_uradial} containing the circular banding noise (associated with \uradial) for the cases of input parameters $f = 0.5$ (corresponding to \uarea) and $f = 1$ (corresponding to \uradial) resulted in filtered images shown in \mfig{img:noisy_img_uradial_filtered_f=0.5} and \mfig{img:noisy_img_uradial_filtered_f=1.0}.

 \subsubsection{Azimuthal banding noise}
\begin{figure}[ht]
	\begin{center}
		\begin{subfigure}[t]{.3\textwidth}
			\centering
			\includegraphics[width=\linewidth]{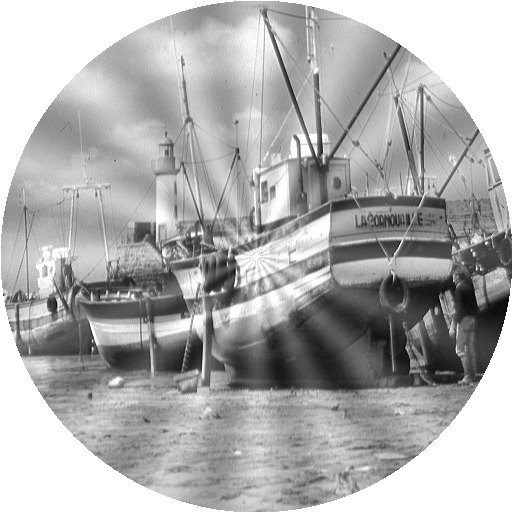}
			\caption{Noisy image.}\label{img:noisy_img_theta}
		\end{subfigure}~~~~~~~~~~~
        \begin{subfigure}[t]{.3\textwidth}
			\centering
			\includegraphics[width=\linewidth]{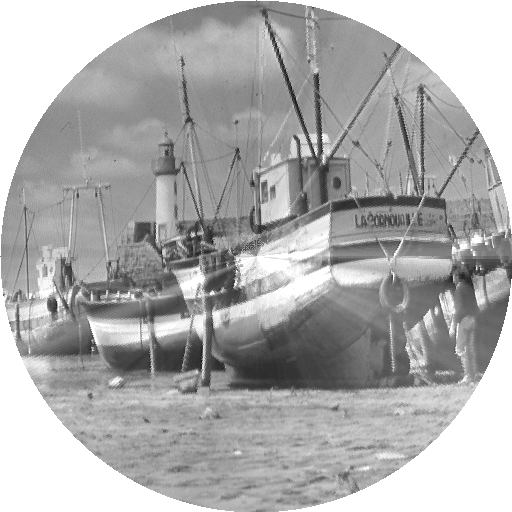}
			\caption{Image filtered with $f$ = 0.5, SSIM = 0.87 ,  PSNR = 23.64, MSE = 1134.64.}\label{img:noisy_img_theta_filtered_f=0.5}
		\end{subfigure}~~~~~~~~~~~
        \begin{subfigure}[t]{.3\textwidth}
			\centering
			\includegraphics[width=\linewidth]{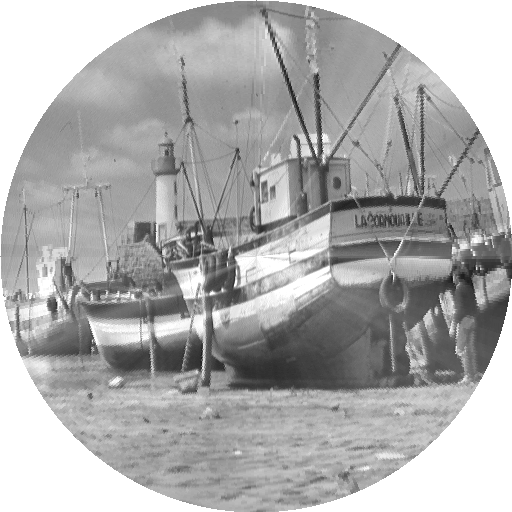}
			\caption{Image filtered with $f$ = 1.0, SSIM = 0.87,  PSNR = 24.97, MSE = 834.60.}\label{img:noisy_img_theta_filtered_f=1.0}
		\end{subfigure}
  \caption{Filtering of the azimuthal banding noise. } \label{img:noisy_azimuthal}
     \end{center}
\end{figure}

An example for illustrating the filtering of the azimuthal banding noise is shown in \mfig{img:noisy_azimuthal}. The noisy image is shown in \mfig{img:noisy_img_theta}. The application of Algorithm \ref{alg_circular_noise_removal} for the cases of input parameters $f = 0.5$ (corresponding to \uarea) and $f = 1$ (corresponding to \uradial) resulted in filtered images. These filtered images are displayed in \mfig{img:noisy_img_theta_filtered_f=0.5} and \mfig{img:noisy_img_theta_filtered_f=1.0} along with image quality metrics such as SSIM, PSNR, and MSE. Since, the noise is not dependent on the radial coordinates,  the image quality metrics SSIM and MSE are observed to be nearly equal for both the cases of input parameters $f = 0.5$ and $f = 1$, as expected.

\subsubsection{Mixed (circular and azimuthal) banding noise}

\begin{figure}[ht]
	\begin{center}
		\begin{subfigure}[t]{.3\textwidth}
			\centering
			\includegraphics[width=\linewidth]{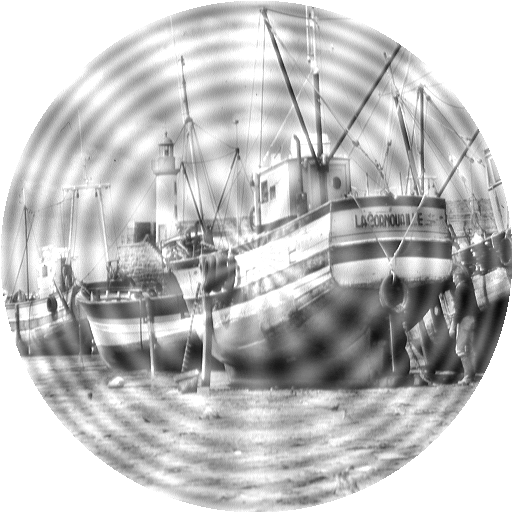}
			\caption{Noisy image.}\label{img:noisy_img_mix+area}
		\end{subfigure}~~~~~~~~~~~
        \begin{subfigure}[t]{.3\textwidth}
			\centering
			\includegraphics[width=\linewidth]{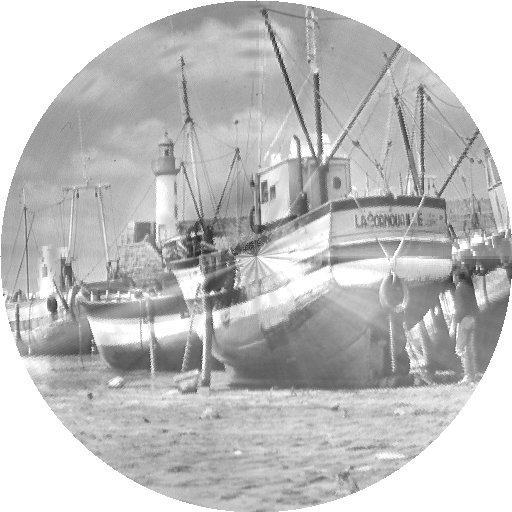}
			\caption{Image filtered with $f$ = 0.5, SSIM = 0.82,  PSNR = 20.00, MSE = 2620.88.}\label{noisy_img_mix_filtered_f=0.5}
		\end{subfigure}~~~~~~~~~~~
        \begin{subfigure}[t]{.3\textwidth}
			\centering
			\includegraphics[width=\linewidth]{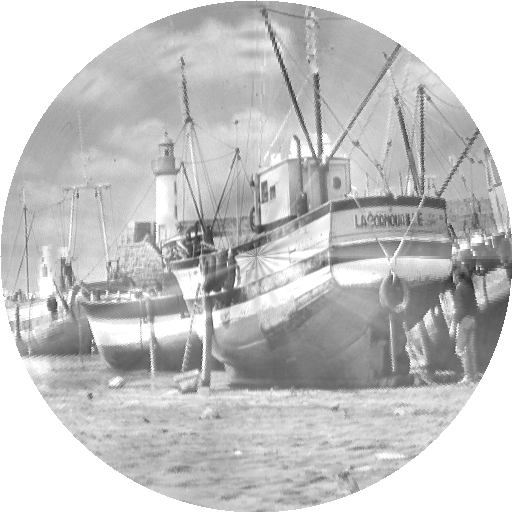}
			\caption{Image filtered with $f$ = 1.0, SSIM = 0.81,  PSNR = 19.78, MSE = 2755.48.}\label{noisy_img_mix_filtered_f=1.0}
		\end{subfigure}
  \caption{Filtering of the combined circular (associated with \uarea) and azimuthal banding noise. }
  \label{img:noisy_mix}
    \end{center}
\end{figure}

We present an example illustrating the filtering of mixed banding noise as shown in \mfig{img:noisy_mix}. In this case, both circular banding noise and azimuthal banding noise are present. The input image affected by noise is shown in \mref{img:noisy_img_mix+area}, while the filtered results obtained using Algorithm~\ref{alg_circular_noise_removal}, with different parameter settings, are displayed in \mfig{noisy_img_mix_filtered_f=0.5} and \mfig{noisy_img_mix_filtered_f=1.0}, corresponding to $f=0.5$ (related to \uarea) and $f=1.0$ (related to \uradial), respectively. Image quality metrics, including SSIM, PSNR, and MSE, are also provided in \mfig{img:noisy_mix}.

\subsubsection{Airy pattern noise}

\begin{figure}[ht]
	\begin{center}
		\begin{subfigure}[t]{.3\textwidth}
			\centering
			\includegraphics[width=\linewidth]{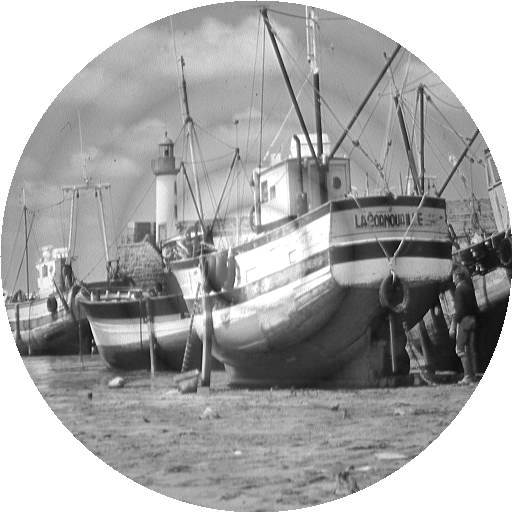}
			\caption{Noisy image.}\label{img:noisy_img_airy}
		\end{subfigure}~~~~~~~~~~~
        \begin{subfigure}[t]{.3\textwidth}
			\centering
			\includegraphics[width=\linewidth]{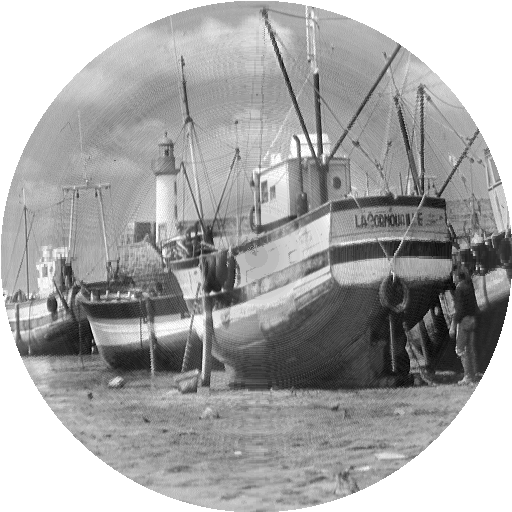}
			\caption{Image filtered with $f$ = 0.5, SSIM = 0.88,  PSNR = 28.81, MSE = 344.60.}\label{img:noisy_img_airy_filtered_f=0.5}
		\end{subfigure}~~~~~~~~~~~
        \begin{subfigure}[t]{.3\textwidth}
			\centering
			\includegraphics[width=\linewidth]{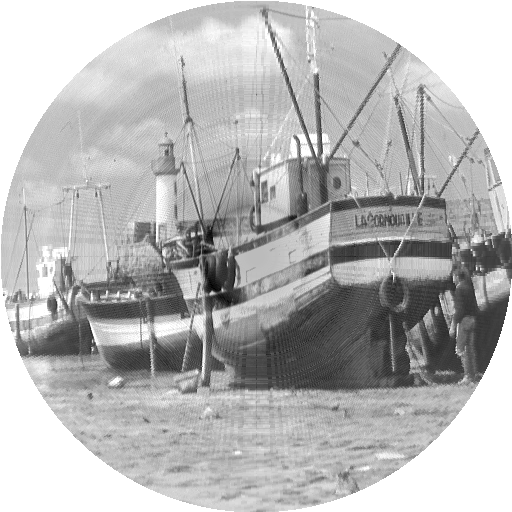}
			\caption{Image filtered with $f$ = 1.0, SSIM = 0.84,  PSNR = 23.34, MSE = 1214.16.}\label{img:noisy_img_airy_filtered_f=1.0}
		\end{subfigure}
  \caption{Filtering of the Airy pattern noise. }
  \label{img:noisy_airy_pattern_example}
    \end{center}
\end{figure}

An example demonstrating the filtering of a noisy image, where the noise has an Airy pattern, is presented in \mfig{img:noisy_airy_pattern_example}. The noisy input image is shown in \mfig{img:noisy_img_airy}, while the filtered images obtained using Algorithm~\ref{alg_circular_noise_removal}, for the cases of input parameters $f = 0.5$ (associated with \uarea) and $f=1.0$ (associated with   \uradial), are shown in \mfig{img:noisy_img_airy_filtered_f=0.5} and \mfig{img:noisy_img_airy_filtered_f=1.0}, respectively.  Further, image quality metrics like SSIM, PNSR and MSE for filtered images are also shown in \mfig{img:noisy_airy_pattern_example}. Mean square error (MSE) is observed to be significantly lower when filtering is based on a \uarea \, (corresponding to $f = 0.5$).

\section{Conclusion}\label{sec:disc}
In this work, we 
presented a novel hybrid classical-quantum approach for image processing based on polar Walsh basis functions.
We introduced a novel hybrid classical-quantum approach for the removal of circular banding noise (including Airy pattern noise) and azimuthal banding noise. This approach is based on a previously established hybrid classical-quantum algorithm for evaluation of Walsh-Hadamard transforms \cite{shukla2021hybrid, shukla2022hybrid}, coupled with the formulation of Walsh basis functions in polar coordinates for image representations presented in this work. 

Our approach presented here provides an innovative solution to the challenges associated with Cartesian to polar transformations of images. We considered 
polar representations based on \uarea \, and \uradial. \Uarea \, provides a polar representation such that each sector ($S(k,q)$ as defined in \meqref{eq:skq}) contains nearly equal number of pixels. One significant aspect addressed in our work is the non-injectivity issue inherent in Cartesian to polar mappings of an image (for example, the sector $S(k,q)$ as defined in \meqref{eq:skq}  may contain more than one pixel or even zero pixels). We devised effective smoothening and interpolating techniques as part of the transformation process, mitigating the challenges posed by the non-injectivity of these mappings.

Since the measurement of a quantum state cannot provide information about the global phase, there are measurement challenges (such as the loss of the sign information) present in extracting useful information. This is also true in the context of the application of the quantum Walsh-Hadamard transforms to row/column vectors of an image. 
The hybrid classical-quantum approach in \cite{shukla2021hybrid, shukla2022hybrid} involving an appropriate adaptation of the quantum Walsh-Hadamard transform provides an efficient method for tackling some of these measurement challenges. We note that the classical Fast Walsh-Hadamard Transform~\cite{beauchamp1975walsh} for an input vector of size \( N \) has a computational complexity of order \( \mathcal{O} (N \log_2 N) \), whereas the hybrid classical-quantum algorithm (Algorithm~\ref{alg_QWHT}, \cite{shukla2021hybrid}) for computation of the Walsh-Hadamard transform for an input vector of size \( N \) is of order \( \mathcal{O}(N) \).
Further, the proposed approach in \cite{shukla2021hybrid} makes efficient use of qubits as it needs only $\log_2 N $ qubits for sequential processing of an image of $ N \times N $ pixels. 
Since Algorithm \ref{alg_circular_noise_removal} presented in this work is based on the previous work in \cite{shukla2021hybrid, shukla2022hybrid}, it inherits all the above-mentioned advantages.

To validate the applicability of our proposed approach, we presented computational examples related to the removal of circular banding noise (including Airy pattern noise) and azimuthal banding noise using Algorithm \ref{alg_circular_noise_removal}. These examples were implemented and successfully tested on the simulation environment on Qiskit (IBM's open source quantum computing platform). 

\section*{{\large Data availability statement}} 
\noindent Data sharing is not applicable to this article as no new data were created or analyzed in this study.

\bibliographystyle{unsrt}

\end{document}